\documentclass[12pt]{article}
\pdfoutput=1
\usepackage{jheppub}

\usepackage{amsmath,amssymb,amsfonts}
\usepackage{graphicx}
\usepackage{color}
\usepackage{tikz}
\usepackage{booktabs}
\usepackage{dsfont}

\usepackage{subfigure}

\def\cA{{\cal A}}
\def\cB{{\cal B}}
\def\cC{{\cal C}}
\def\cF{{\cal F}}
\def\cG{{\cal G}}
\def\cJ{{\cal J}}
\def\cY{{\cal Y}}
\def\cK{{\cal K}}
\def\cL{{\cal L}}
\def\cS{{\cal S}}
\def\mf{{\mathfrak F}}
\def\ms{{\mathfrak s}}

\def\n{{\mathfrak n}}

\DeclareMathOperator{\vol}{vol}
\DeclareMathOperator{\Vol}{Vol}

\DeclareMathOperator{\Res}{Res}
\newcommand{\bea}{\begin{eqnarray}}
\newcommand{\eea}{\end{eqnarray}}

\makeatletter\def\l@subsubsection#1#2{}%
\makeatother

\def\Im{\mathop{\rm Im}}
\def\Re{\mathop{\rm Re}}
\def\RR{\mathds{R}}

\def\CC{\mathds{C}}

\begin{document}

\title{Type IIB 7-branes in warped \texorpdfstring{$AdS_6$}{AdS6}: partition functions, brane webs and probe limit}

\author{Michael Gutperle, }
\emailAdd{gutperle@physics.ucla.edu}
\author{Andrea Trivella, }
\emailAdd{andrea.trivella@physics.ucla.edu}
\author{Christoph F.~Uhlemann} 
\emailAdd{uhlemann@physics.ucla.edu}

\affiliation{Mani L.\ Bhaumik Institute for Theoretical Physics\\
Department of Physics and Astronomy\\
University of California, Los Angeles, CA 90095, USA}

\abstract{
We study Type IIB supergravity solutions with spacetime of the form $AdS_6\times S^2$ warped over a Riemann surface $\Sigma$,
where $\Sigma$ includes punctures around which the supergravity fields have non-trivial $SL(2,\RR)$ monodromy. Solutions without monodromy have a compelling interpretation as near-horizon limits of $(p,q)$ 5-brane webs, and the punctures have been interpreted as additional 7-branes in the web.
In this work we provide further support for this interpretation and clarify several aspects of the identification of the supergravity solutions with brane webs.
To further support the identification of the punctures with 7-branes, we show that punctures with infinitesimal monodromy match a probe 7-brane analysis using $\kappa$-symmetry. We then construct families of solutions with fixed 5-brane charges and punctures with finite monodromy, corresponding to fully backreacted 7-branes. We compute the sphere partition functions of the dual 5d SCFTs and use the results to discuss concrete brane web interpretations of the supergravity solutions.}

\maketitle

\section{Introduction and summary}

Five-dimensional superconformal field theories (SCFTs) have been studied extensively since first concrete evidence for their existence has been presented in \cite{Seiberg:1996bd,Intriligator:1997pq}. They exhibit many interesting phenomena, not the least of which is that they can not be treated consistenly in conventional perturbative quantization schemes. This makes indirect methods, such as their engineering in string theory and the AdS/CFT dualities, particularly valuable.
Large classes of 5d SCFTs can indeed be engineered in Type IIB string theory via $(p,q)$ 5-brane webs \cite{Aharony:1997ju,Aharony:1997bh}, which describe gauge theory deformations of the 5d SCFTs and in the limit where the web collapses to a 5-brane intersection at a point, describe the SCFT itself. In \cite{DHoker:2016ujz,DHoker:2016ysh,DHoker:2017mds}\footnote{Earlier work in the context of Type IIB can be found in \cite{Lozano:2012au,Lozano:2013oma,Apruzzi:2014qva,Kim:2015hya,Kim:2016rhs}, while solutions in Type IIA have been discussed in \cite{Brandhuber:1999np,Bergman:2012kr}.} supergravity solutions were constructed which are in one-to-one correspondence with 5-brane intersections and which provide compelling candidates for holographic duals to the SCFTs realized on such intersections. This allows to use the established tools of AdS/CFT for quantitative analyses of the 5d SCFTs.

The space of 5d SCFTs that can be realized in Type IIB string theory can be extended substantially by adding additional 7-branes into 5-brane webs \cite{DeWolfe:1999hj}, and many insights have been obtained through the inclusion of 7-branes and in particular their associated branch cuts \cite{Benini:2009gi,Taki:2014pba,Bergman:2014kza,Kim:2015jba,Hayashi:2015fsa,Hayashi:2015zka}. This motivates a corresponding extension of the construction of supergravity solutions. In \cite{DHoker:2017zwj} the construction of supergravity solutions has indeed been extended to incorporate punctures with non-trivial $SL(2,\RR)$ monodromy, signaling the presence of additional 7-branes. However, while the map between supergravity solutions and 5-brane webs appeared very clearly and naturally in the case without monodromy, where a given 5-brane intersection is entirely characterized by the charges of the external 5-branes, a corresponding map is less automatic in the case with additional 7-branes. This is largely due to the fact that 7-branes introduce a number of additional parameters, as we will review shortly and in more detail in sec.~\ref{sec:review}, and the fact that the analysis of the supergravity solutions is technically more challenging. This motivates further study of the solutions with monodromy, to substantiate and clarify their interpretation.

The solutions in \cite{DHoker:2017zwj} are constructed in terms of two locally holomorphic functions $\cA_\pm$ on the Riemann surface $\Sigma$, which is a disc or equivalently the upper half plane. The differentials of these functions have common poles on the boundary of $\Sigma$, at which the entire solution approaches that for a $(p,q)$ 5-brane, as constructed in \cite{Lu:1998vh}, with $p-iq$ identified with the residue at the pole. This facilitates the identification of the solutions with $(p,q)$ 5-brane webs. For solutions with monodromy, the differentials in addition have a number of branch points in the interior of $\Sigma$ with associated branch cuts, across which the supergravity fields undergo a parabolic $SL(2,\RR)$ transformation. The regularity conditions for the supergravity solutions as constructed in \cite{DHoker:2017zwj} constrain each puncture to lie on a curve in $\Sigma$. This leaves one real parameter in addition to the orientation of the branch cut for a puncture with fixed monodromy. Adding a 7-brane into a 5-brane web correspondingly adds new parameters. In addition to the orientation of the branch cut, there is a choice of which face of the web the 7-brane is placed in. This choice remains meaningful in the conformal limit and naturally turns into a continuous parameter in a ``large-$N$'' limit, thus providing a potential brane web realization of the supergravity parameter. One may wonder, however, whether a given puncture corresponds to an isolated 7-brane in a certain face of the web, or whether 5-branes are attached to it. Similarly, one may wonder whether solutions with punctures at different points in $\Sigma$ can be related by 7-brane moves with the associated Hanany-Witten brane creation effect \cite{Hanany:1996ie}, or whether punctures at different points correspond to genuinely different brane webs. An unambiguous brane web interpretation for the solutions constructed in \cite{DHoker:2017zwj} is therefore not immediately clear. In the present paper we will expand on the interpretation of the solutions in \cite{DHoker:2017zwj} in several ways and address these questions. We will constrain the monodromy around the punctures to realize the $SL(2,\RR)$ transformation appropriate for D7 branes for simplicity, but the results immediately generalize to other 7-branes by globally conjugating with suitable $SL(2,\RR)$ elements.

As a first step we will provide further support for the identification of the punctures with 7-branes in sec.~\ref{sec:kappa}, by connecting the solutions with punctures and $SL(2,\RR)$ monodromy to a probe brane analysis. The strength of the $SL(2,\RR)$ monodromy around a given puncture is given in a precise way by the 7-brane charge at the puncture, and in the limit where the monodromy transformation becomes infinitesimally close to the identity, we expect to recover a solution without puncture but with an additional probe D7 brane embedded into it. We will show that this is indeed the case. 
We will study warped $AdS_6$ solutions without punctures, and derive the BPS equations for supersymmetric probe D7 branes wrapping $AdS_6\times S^2$ in these solutions. The probe BPS equations are derived from a $\kappa$-symmetry analysis, and we will clarify an important subtlety in this analysis which arises due to the presence of non-trivial axion-dilaton backgrounds: The $\kappa$-symmetry conditions of \cite{Bergshoeff:1996tu, Cederwall:1996pv,Cederwall:1996ri} are derived with a particular gauge fixing of the local $U(1)$ in the covariant formulation of the Type IIB supergravity field equations of \cite{Schwarz:1983qr,Howe:1983sra}. The analysis of the supergravity BPS equations in \cite{DHoker:2016ujz,DHoker:2016ysh,DHoker:2017mds,DHoker:2017zwj}, on the other hand, was carried out with a different gauge fixing. This has to be accounted for when using the expressions for the Killing spinors of the warped $AdS_6$ solutions in the $\kappa$-symmetry conditions, as we will explain in detail in sec.~\ref{sec:su11}. Once this subtlety is taken into account, we find that the BPS equations for a probe D7 brane in a solution without monodromy, which constrain its location in $\Sigma$, precisely reproduce the regularity conditions for a supergravity solution with puncture in the limit in which the monodromy is infinitesimally close to the identity. This shows that the puncture, in the probe limit, can indeed be identified with a probe D7 brane.

We will then turn to the solutions with punctures and finite monodromy, corresponding to fully backreacted 7-branes, in sec.~\ref{sec:backreacted}. A crucial point for a subsequent brane web interpretation of our results is that we will consider families of solutions where the physical 5-brane charges are fixed, while the location of the puncture, the orientation of the branch cut and the 7-brane charge are allowed to vary. 
We will in particular study the $S^5$ partition functions of the dual SCFTs, which can be conveniently extracted holographically from the minimal surface computing the entanglement entropy of a ball-shaped region \cite{Casini:2011kv}. The partition functions are expected to agree for supergravity solutions describing brane webs that realize the same SCFT, and therefore provide crucial information for understanding the brane web interpretation.
In sec.~\ref{sec:branch-cut-dependence} we will show that the partition functions are generally invariant under changes in the orientation of a branch cut, provided that no poles are crossed. This is consistent with the interpretation that solutions with the same 5- and 7-brane charges that differ only in the orientation of the branch cut describe the same dual SCFT, as one would expect from the brane web picture. 

In sec.~\ref{sec:D5-NS5x2-D7}, we will realize a family of solutions with 3 poles, two corresponding to NS5 branes and one corresponding to D5 branes, and one puncture corresponding to D7 branes. Such a solution would not be possible without monodromy since the 5-brane charges do not add to zero, but they can be realized in the presence of D7 branes. The regularity conditions constrain the D7-brane puncture to lie on a curve in $\Sigma$ which starts at the D5 brane pole and ends between the two NS5 brane poles. Fixing the precise form of the 5-brane charges, independently from the position of the puncture, leads to a non-trivial relation between the charge of the 7-brane and the position of the puncture. This shows that configurations with the same 5-brane charges but punctures at different locations are not related by Hanany-Witten transitions. The 7-brane charge that is required to realize a given set of 5-brane charges increases as the puncture approaches the boundary of $\Sigma$, and we show that, as the puncture is moved onto the boundary, the solution reduces to a 4-pole solution without monodromy, where the puncture with diverging charge produces the additional pole with the appropriate 5-brane charge on the boundary of $\Sigma$. The computation of the partition function, which can be given analytically up to a single function of one parameter that we provide numerically, shows that it has a non-trivial dependence on the position of the puncture on $\Sigma$. This further shows that solutions with the same 5-brane charges but a puncture at different locations on $\Sigma$ realize genuinely different dual SCFTs. As the puncture is moved to the boundary of the disc, the partition function approaches that of a 4-pole solution without monodromy, as expected from the limiting procedure discussed above.

In sec.~\ref{sec:4-pole} we will realize a family of 4-pole solutions with two NS5 poles, two D5 poles and one D7-brane puncture. The D5 brane charges do not sum to zero and such solutions could again not be realized without 7-branes. An interesting feature of these solutions is that the branch cut associated with the puncture intersects the boundary directly on one of the D5-brane poles. This has a natural interpretation in the brane web picture as a branch cut going out to infinity within a stack of external D5 branes. The puncture can be placed on a curve in $\Sigma$ that connects the two D5-brane poles, and the D7-brane charge that is required to realize a given set of 5-brane charges now depends on the difference in D5-brane charge between the two D5-brane poles as well as on the location of the puncture. The partition function can be given analytically up to two functions, each of one parameter, that we provide numerically.
The family of 3-pole solutions with one puncture discussed in sec.~\ref{sec:D5-NS5x2-D7} can be obtained as a special case from this family of 4-pole solutions, where the residue at one D5-brane pole vanishes. Consistency of this limit imposes a relation between the partition functions for the 3- and 4-pole solutions with puncture, and we indeed find this relation to be satisfied. In general, the partition function in the four-pole solution again is a non-trivial function of the location of the puncture for fixed 5-brane charges, showing again that solutions with punctures at different points in $\Sigma$ describe different SCFTs.

Finally, in sec.~\ref{sec:brane-web} we will discuss the families of 3- and 4-pole solutions with fixed 5-brane charges and punctures in the context of a brane web interpretation for the supergravity solutions. We will use the dependence of the 7-brane charge on the location of the puncture, the results on the partition functions, and the limiting procedures relating the various families of solutions to clarify the identification of the supergravity solutions with brane webs, and devise a consistent picture for their interpretation.

\subsection{Outline}
The remainder of the paper is organized as follows. In sec.~\ref{sec:review} we review the warped $AdS_6\times S^2\times\Sigma$ solutions with and without monodromy. In sec.~\ref{sec:kappa} we study supersymmetric probe D7 brane emeddings into the solutions without monodromy, and compare to the solutions with monodromy.
In sec.~\ref{sec:backreacted} we turn to the solutions with fully backreacted 7-branes, discuss the entanglement entropy of a ball-shaped region from which the $S^5$ partition function can be extracted, construct families of solutions with fixed 5-brane charges and explicitly compute the partition functions. In sec.~\ref{sec:brane-web} we discuss the brane interpretation of the results.
In the appendices we derive the regularity conditions for the case where one pole is at the point at infinity of the upper half plane, discuss the relation of the BPS and the field equations for probe D7 branes, and provide explicit expressions for the background 7- and 9-form field strengths.

\section{Review of warped \texorpdfstring{$AdS_6\times S^2\times \Sigma$}{AdS6xS2xSigma} solutions}\label{sec:review}
To fix notation we will provide a brief review of the warped $AdS_6\times S^2\times\Sigma$ solutions to type IIB supergravity without monodromy constructed in \cite{DHoker:2016ysh,DHoker:2017mds}, and of the extension to incorporate punctures provided in \cite{DHoker:2017zwj}.
The non-vanishing fields of type IIB supergravity in the conventions of \cite{Schwarz:1983qr,Howe:1983sra} are the metric, the axion-dilaton scalar $B$ and the complex two-form $\hat C_{(2)}$, where we introduced a hat to avoid confusion with the real R-R potential $C_{(2)}$.
With a complex coordinate $w$ on $\Sigma$, which is taken to be the upper half plane, the metric and the 2-form field are parametrized by scalar functions $f_2^2$, $f_6^2$, $\rho^2$ and $\cC$ on $\Sigma$ as follows,
\begin{align}\label{eqn:ansatz}
 ds^2 &= f_6^2 \, ds^2 _{AdS_6} + f_2 ^2 \, ds^2 _{S^2} + 4\rho^2 dw d\bar w~,
 &
 \hat C_{(2)}&=\cC \vol_{S^2}~,
\end{align}
where $\vol_{S^2}$ is the volume form on $S^2$ of unit radius.
The solutions are expressed in terms of two locally holomorphic functions $\cA_\pm$ and the following composite quantities
\begin{align}\label{eq:kappa-G}
 \kappa^2&=-|\partial_w \cA_+|^2+|\partial_w \cA_-|^2~,
 &
 \partial_w\cB&=\cA_+\partial_w \cA_- - \cA_-\partial_w\cA_+~,
 \\
 \cG&=|\cA_+|^2-|\cA_-|^2+\cB+\bar{\cB}~,
 &
  R+\frac{1}{R}&=2+6\,\frac{\kappa^2 \, \cG }{|\partial_w\cG|^2}~.
  \label{eqn:Gdef}
\end{align}
The explicit form of the functions parametrizing the metric is then given by
\begin{align}\label{eqn:metric}
f_6^2&=\sqrt{6\cG} \left ( \frac{1+R}{1-R} \right ) ^{1/2},
&
f_2^2&=\frac{1}{9}\sqrt{6\cG} \left ( \frac{1-R}{1+R} \right ) ^{3/2},
&
\rho^2&=\frac{\kappa^2}{\sqrt{6\cG}} \left (\frac{1+R}{1-R} \right ) ^{1/2},
\end{align}
where we used the expressions of \cite{DHoker:2017mds} with $c_6=1$.
The function $\cC$ parametrizing the complex 2-form field is given by
\begin{align}\label{eqn:flux}
 \cC = \frac{4 i }{9}\left (  
\frac{\partial_{\bar w} \bar \cA_- \, \partial_w \cG}{\kappa ^2} 
- 2 R \, \frac{  \partial_w \cG \, \partial_{\bar w} \bar \cA_- +  \partial_{\bar w}  \cG \, \partial_w \cA_+}{(R+1)^2 \, \kappa^2 }  
 - \bar  \cA_- - 2 \cA_+ \right )~,
\end{align}
and the axion-dilaton scalar $B$ is given by 
\begin{align}\label{eq:B}
B &=\frac{\partial_w \cA_+ \,  \partial_{\bar w} \cG - R \, \partial_{\bar w} \bar \cA_-   \partial_w \cG}{
R \, \partial_{\bar w}  \bar \cA_+ \partial_w \cG - \partial_w \cA_- \partial_{\bar w}  \cG}~.
\end{align}
These configurations solve the BPS equations for preserving sixteen supersymmetries, and as shown in \cite{Corbino:2017tfl} also the equations of motion.
A crucial ingredient for the $\kappa$-symmetry analysis will be the form of the Killing spinors. We will use the Clifford algebra conventions summarized in appendix A of \cite{DHoker:2016ujz}. The ten-dimensional Killing spinor $\epsilon$ is expanded in terms of $AdS_6\times S^2$ Killing spinors $\chi^{\eta_1\eta_2}$ and complex two-component spinors on $\Sigma$, $\zeta_{\eta_1\eta_2}$, as follows
\begin{align}
 \epsilon&=\sum_{\eta_1,\eta_2=\pm}\chi^{\eta_1\eta_2}\otimes\zeta_{\eta_1\eta_2}~,
\end{align}
and analogously\footnote{To avoid confusion with the composite quantity $\cB$ defined in (\ref{eq:kappa-G}), we will denote the charge conjugation matrix by $C$ throughout.} $C^{-1}\epsilon^\star=\sum_{\eta_1\eta_2}\chi^{\eta_1\eta_2}\otimes \star\zeta_{\eta_1\eta_2}$, with $\star\zeta_{\eta_1\eta_2}=-i\eta_2\sigma^2\zeta^\star_{\eta_1\,-\eta_2}$.
In a chirality basis where $\sigma^3$ is diagonal, we have
\begin{align}\label{eq:zeta}
 \zeta_{++}&=\begin{pmatrix}   \bar\alpha\\ \beta     \end{pmatrix}~,
 &
 \zeta_{--}&=\begin{pmatrix}   -\bar\alpha\\ \beta     \end{pmatrix}~,
  &
 \zeta_{+-}&=i\nu\zeta_{++}~,
 &
 \zeta_{-+}&=i\nu\zeta_{--}~.
\end{align}
where $\nu\in\lbrace-1,+1\rbrace$ and, with $f^{-2}=1-|B|^2$,
\begin{align}\label{eq:alphabeta}
 \rho\bar\alpha^2&=f(\partial_w\cA_+ + B \partial_w\cA_-)~,
 &
 \rho\beta^2&=f(B\partial_{\bar w}\bar\cA_+ +\partial_{\bar w}\bar\cA_-)~.
\end{align}
The action of the Clifford algebra elements on the Killing spinors that will be relevant for the discussion of $\kappa$-symmetry are derived from the relation
\begin{align}
 (\gamma_{(1)}\otimes I_2)\chi^{\eta_1\eta_2}&=\chi^{-\eta_1\eta_2}~,
 &
 (I_8\otimes \gamma_{(2)})\chi^{\eta_1\eta_2}&=\chi^{\eta_1\,-\eta_2}~,
\end{align}
where $\gamma_{(i)}$ denotes the chirality matrices on the respective components of $AdS_6\times S^2\times\Sigma$ (see appendix A of \cite{DHoker:2016ujz} for more details).
From these one concludes that
\begin{align}
 \Gamma^{01234567}\epsilon&=-i\sum_{\eta_1\eta_2}\chi^{\eta_1\eta_2}\otimes\zeta_{-\eta_1\,-\eta_2}~,
 \nonumber\\
 \Gamma^{67}\Gamma^{01234567}C^{-1}\epsilon^\star&=\sum_{\eta_1\eta_2}\chi^{\eta_1\eta_2}\otimes \star\zeta_{-\eta_1\eta_2}~.
 \label{eq:Gamma-eps}
\end{align}

\subsection{Solutions without monodromy}\label{sec:sol-no-d7}
The physically regular solutions without monodromy constructed in \cite{DHoker:2016ysh,DHoker:2017mds} amount to a particular choice of the locally holomorphic functions $\cA_\pm$ on the upper half plane, which is given by
\begin{align}\label{eqn:cA-0}
 \cA_\pm (w) &=\cA_\pm^0+\sum_{\ell=1}^L Z_\pm^\ell \ln(w-p_\ell)~,
\end{align}
where the $p_\ell$ are poles with residues $Z_\pm^\ell$ in $\partial_w \cA_\pm$, that are restricted to be on the real line. 
The constants $\cA_\pm^0$ are constrained by $\bar\cA_\pm^0=-\cA_\mp^0$.
The residues are given in terms of complex parameters $s_n$ that are constrained to lie in the interior of $\Sigma$ as follows,
\begin{align}\label{eqn:residues}
Z_+^\ell  &=
 \sigma\prod_{n=1}^{L-2}(p_\ell-s_n)\prod_{k \neq\ell}^L\frac{1}{p_\ell-p_k}~,
 &
 Z_-^\ell&= - \overline{Z_+^\ell}~.
\end{align}
For these solutions to satisfy the desired regularity conditions, the parameters appearing in the locally holomorphic functions have to be constrained to satisfy
\begin{align}
\label{eqn:constr}
 \cA_+^0 Z_-^k - \cA_-^0 Z_+^k 
+ \sum _{\ell \not= k }Z^{[\ell k]} \ln |p_\ell - p_k| &=0~,
\end{align}
where $Z^{[\ell k]}\equiv Z_+^\ell Z_-^k-Z_+^k Z_-^\ell$.
The crucial feature for the identification of the solutions with 5-brane webs is that the $2L-2$ free parameters of a solution with $L$ poles can be taken as the residues $Z_+^\ell$, subject to the constraint that $\sum_\ell Z_+^\ell=0$.
Combined with the observation that at each pole $p_m$ the solution turns into a $(q_1,q_2)Q$ 5-brane solution, in the conventions of \cite{Lu:1998vh}, with
\begin{align}\label{eq:5-brane-charge}
 (q_1-iq_2)Q&=\frac{8}{3}Z_+^m~,
\end{align}
this gives a direct identification of the supergravity solutions with 5-brane intersections.

\subsection{Solutions with monodromy}\label{sec:sol-d7}
We will now briefly review the construction to add punctures with monodromy to the solutions without monodromy summarized above.
We will exclusively focus on punctures with D7-brane monodromy in this paper, and refer to \cite{DHoker:2017zwj} for the more general case.
Note, however, that with no restrictions on the residues at the poles on $\partial\Sigma$, the case of punctures with generic (commuting) parabolic $SL(2,\RR)$ monodromies can be obtained straightforwardly from the results presented here by global $SL(2,\RR)$ transformations. In that sense the restriction to D7-brane monodromy is without loss of generality.

In addition to the parameters for the solutions without monodromy, a solution with D7-brane punctures depends on the loci of the punctures, $w_i$, $i=1,\dots,I$, a real number $n_i$ for each puncture and a phase $\gamma_i$ specifying the orientation of the branch cut. From this data one constructs a function $f$, which encodes the branch points and branch cut structure, via
\begin{align}
f(w) = \sum _{i=1}^I \frac{n_i^2}{4\pi} \ln \left ( \gamma_i\,\frac{ w-w_i}{w -\bar w_i} \right )~.
\end{align}
With the help of this function and $Y^\ell\equiv Z_+^\ell-Z_-^\ell$, the locally holomorphic functions for a solution with monodromy are expressed as
\begin{align}\label{eqn:cA-monodromy}
 \cA_\pm&= \cA_\pm^0+\sum_{\ell=1}^L Z_\pm^\ell \ln(w-p_\ell) + \int_\infty^w dz \;f(z)\sum_{\ell=1}^L \frac{Y^\ell}{z-p_\ell}~,
\end{align}
again with $\bar\cA_\pm^0=-\cA_\mp^0$. The contour for the integration is chosen such that it does not cross any of the branch cuts.
The regularity constraints that the parameters have to satisfy for the solutions with D7-brane monodromy are
\begin{align}
\label{eq:w1-summary}
 0&=2\cA_+^0-2\cA_-^0+\sum_{\ell=1}^LY^\ell \ln|w_i-p_\ell|^2~,
 &i&=1,\cdots,I~,
 \\
 0&=2\cA_+^0\cY_-^k-2\cA_-^0\cY_+^k
 +\sum_{\ell\neq k} Z^{[\ell, k]}\ln |p_\ell-p_k|^2+Y^kJ_k~,
 &
 k&=1,\cdots,L~.
 \label{eq:DeltaG0-summary}
\end{align}
With $\cS_k\subset\lbrace 1,\cdots, I\rbrace$ denoting the set of branch points for which the associated branch cut intersects the real line in the interval $(p_k,\infty)$, $J_k$ is given by
\begin{align}\label{eq:Jk}
 J_k&=\sum_{\ell=1}^L Y^\ell\Bigg[\int_\infty^{p_k} dx f^\prime(x)  \ln |x-p_\ell|^2
 +\sum_{i\in\cS_k} \frac{i n_i^2}{2} \ln |w_i-p_\ell|^2\Bigg]~.
\end{align}
The residues of the differentials of (\ref{eqn:cA-monodromy}) at the poles are given by
\begin{align}\label{eq:cY}
\cY_\pm^\ell&=Z_\pm^\ell+ f(p_\ell)Y^\ell~.
\end{align}
It is these residues that translate to the charges of the external 5-branes and replace the $Z_+^\ell$ in (\ref{eq:5-brane-charge}), resulting in
\begin{align}\label{eq:5-brane-charge-Y}
 (q_1-iq_2)Q&=\frac{8}{3}\cY_+^m~.
\end{align}

\section{Match to probe D7 branes and \texorpdfstring{$\kappa$}{kappa}-symmetry}\label{sec:kappa}

In this section we study probe D7 branes embedded into the solutions reviewed in sec.~\ref{sec:review}, subject to the requirement that they preserve all bosonic and fermionic symmetries of the background. This is motivated by the fact that the solutions with and without punctures discussed in sec.~\ref{sec:review} are both invariant under $SO(2,5)\oplus SO(3)$ and sixteen supersymmetries.
The requirement to preserve the bosonic symmetries forces the D7-branes to wrap the entire $AdS_6\times S^2$ part of the geometry, and the entire embedding is therefore characterized by the point at which the D7-branes are localized in $\Sigma$. 
The choice of coordinates on $AdS_6$ is irrelevant for the analysis, and we will therefore leave it general.
The worldvolume metric induced by the string-frame background metric on the D7-brane reads
\begin{align}\label{eq:D7-metric}
 g&=\tilde f_6(w,\bar w)^2 ds^2_{AdS_6}+\tilde f_2(w,wb)^2 ds^2_{S^2}~,
\end{align}
where the tilde denotes that the radii are in string frame. The pullback of the ten-dimensional frame to the D7-brane, $E^a$, is given by
\begin{align}
 E^m&=\tilde f_6 \hat e^m~, & m&=0,\dots,5~,
 \nonumber\\
 E^i&=\tilde f_2 \hat e^i~, & i&=6,7~,
 \nonumber\\
 E^8=E^9&=0~,
\end{align}
where $\hat e^m$ and $\hat e^i$ denote the canonical frames for $AdS_6$ and $S^2$, respectively.
The symmetry requirement constrains the field strength of the worldvolume gauge field, $F$, to be proportional to the volume form of $S^2$, and we can thus parametrize it as
\begin{align}\label{eq:F-0}
 F&=\cK \vol_{S^2}~,
\end{align}
where $\vol_{S^2}$ is the canonical volume form on $S^2$ of unit radius and $\cK$ is a real constant to be solved for for each supersymmetric embedding.

\subsection{\texorpdfstring{$\kappa$}{kappa}-symmetry and \texorpdfstring{$SU(1,1)/U(1)$}{SU(1,1)/U(1)}}\label{sec:su11}
The supersymmetries preserved by a probe brane embedding are those generated by background Killing spinors $\epsilon$ that are compatible with the $\kappa$-symmetry condition
\begin{align}\label{eqn:kappa-00}
 \Gamma_\kappa\epsilon&=\epsilon~,
\end{align}
where $\Gamma_\kappa$ is a projector that depends on the embedding and has been constructed in \cite{Bergshoeff:1996tu, Cederwall:1996pv,Cederwall:1996ri}. 
The condition will provide constraints on the background fields, that single out the locations where probe branes can be added while preserving supersymmetry.
The explicit expression for $\Gamma_\kappa$ is given by
\begin{align}\label{eqn:kappa-1}
  \Gamma_\kappa&=\frac{1}{\sqrt{\det(1+X)}}\sum_{n=0}^\infty\frac{1}{2^n n!}\gamma^{j_1k_1\dots j_nk_n}X_{j_1k_1}\dots X_{j_nk_n}J_{(p)}^{(n)}~,
\end{align}
where the $\gamma_\mu\equiv E_\mu^a\Gamma_a$ are the pullback of the background Clifford algebra generators to the 7-brane worldvolume, $X^i_{\hphantom{i}j}\equiv g^{ik}\mathcal F_{kj}$, $g$ is the metric induced on the worldvolume by the string-frame background metric, and $\mathcal F$ is defined in terms of the worldvolume field strength $F$ and the background NS-NS two-form field $B_2$ as
\begin{align}\label{eq:cF-def}
 \mathcal F&=F-B_2~.
\end{align}
For $J_{(p)}^{(n)}$ we will use the conventions for complex spinors as spelled out in sec.~2.2 of \cite{Karch:2015vra}, such that
\begin{align}
 J_{(p)}^{(n)}\epsilon&=i(-1)^{(p-1)/2} 
 \begin{cases}
                          \Gamma_{(0)}\epsilon & n+(p-3)/2 \text{\ even}
                          \\
                          C\left(\Gamma_{(0)}\epsilon\right)^\star & n+(p-3)/2 \text{\ odd}
                         \end{cases}~,
\end{align}
with $\Gamma_{(0)}$ given by
\begin{align}\label{eq:Gamma0}
  \Gamma_{(0)}&=\frac{1}{(p+1)!\sqrt{-\det g}}\,\varepsilon^{i_1\dots i_{p+1}}\gamma^{}_{i_1\dots i_{p+1}}~. 
\end{align}
We note in particular that $\Gamma_\kappa$ is not a $\CC$-linear operator, which will play a role shortly.

A crucial subtlety in the formulation of the $\kappa$-symmetry conditions in the backgrounds we are interested in arises due to the presence of non-trival axion-dilaton backgrounds. The $\kappa$-symmetry conditions derived in \cite{Bergshoeff:1996tu, Cederwall:1996pv,Cederwall:1996ri} and the supergravity solutions in \cite{DHoker:2016ujz,DHoker:2016ysh,DHoker:2017mds,DHoker:2017zwj} are both formulated in terms of the physical axion and dilaton fields. This amounts to passing from the formulation of type IIB supergravity in \cite{Schwarz:1983qr,Howe:1983sra}, with linear $SU(1,1)$ action and $U(1)$ gauge symmetry, to gauge-fixed versions. In the notation used in sec.~2 of \cite{DHoker:2016ujz}, the covariant formulation in particular involves a complex one-form $P$, which is constrained by Bianchi identities and transforms under the $U(1)$ as
\begin{align}
 P&\rightarrow e^{2i\theta}P~.
\end{align}
Crucially for the $\kappa$-symmetry analysis, the generators of (local) supersymmetries transform under this $U(1)$ as
\begin{align}\label{eq:eps-U1}
 \epsilon&\rightarrow e^{i\theta/2}\epsilon~.
\end{align}
Expressing $P$ and $Q$ in terms of physical fields was done in  \cite{DHoker:2016ujz} by the following choice for $P$
\begin{align}
 P&=\frac{dB}{1-|B|^2}~,&
 B&=\frac{1+i\tau}{1-i\tau}~.
\end{align}
In contrast, as discussed in sec.~3 of \cite{Cederwall:1996pv}, the expression used for the derivation of the $\kappa$-symmetry condition is
\begin{align}
 P_\kappa&=\frac{d\tau}{\bar\tau-\tau}~.
\end{align}
These two choices are related by a $U(1)$ transformation as follows
\begin{align}
 P&= e^{2i\theta_\kappa}P_\kappa ~,
 &
 e^{2i\theta_\kappa}&=\frac{1+i\bar\tau}{1-i\tau}~.
\end{align}
Consequently, the background Killing spinors used in the $\kappa$ symmetry condition have to be transformed according to (\ref{eq:eps-U1}) to get the condition in the conventions used for the supergravity solutions. Since $\Gamma_\kappa$ is in general not a $\CC$-linear operator, this modifies the condition in a non-trivial way.
We multiply (\ref{eqn:kappa-00}) by $e^{i\theta_\kappa/2}$, and may then state the converted condition as follows:
The supersymmetries preserved by a probe brane embedding in the solutions of \cite{DHoker:2016ujz,DHoker:2016ysh,DHoker:2017mds,DHoker:2017zwj} are those generated by Killing spinors compatible with
\begin{align}\label{eqn:kappa-0}
 \Gamma_\kappa\epsilon&=\epsilon~,
\end{align}
where $\Gamma_\kappa$ is as given in (\ref{eqn:kappa-1}) and
\begin{align}\label{eq:J-new}
 J_{(p)}^{(n)}\epsilon&=i(-1)^{(p-1)/2} 
 \begin{cases}
                          \Gamma_{(0)}\epsilon & n+(p-3)/2 \text{\ even}
                          \\
                          e^{i\theta_\kappa}C\left(\Gamma_{(0)}\epsilon\right)^\star & n+(p-3)/2 \text{\ odd}
                         \end{cases}~,
\end{align}
with $\Gamma_{(0)}$ as given in (\ref{eq:Gamma0}) and $\epsilon$ in $\ref{eqn:kappa-0}$ referring to spinors in the supergravity conventions of \cite{DHoker:2016ujz,DHoker:2016ysh,DHoker:2017mds,DHoker:2017zwj}. We note that the phase $e^{i\theta_\kappa}$ occured for similar reasons in the (re)definition of the three-form field in \cite{Grana:2001xn}.

\subsection{BPS equations for D7-branes}
We now turn to the specific case of probe D7 branes wrapping $AdS_6\times S^2$. 
We identify the NS-NS two-form field $B_2$ and the R-R two-form potential $C_{(2)}$ with the real and imaginary parts of the complex two-form parametrized by $\cC$ as follows,
\begin{align}\label{eq:B2-C2-cC}
 B_2+i C_{(2)}&=\cC\vol_{S^2}~.
\end{align}
With the form of $F$ in (\ref{eq:F-0}) we then have
\begin{align}\label{mf-def}
 \mathcal F&=\mf\vol_{S^2}~,
 &
 \mf&=\cK-\Re(\cC)~.
\end{align}
The sum in (\ref{eqn:kappa-1}) therefore terminates at $n=1$. From (\ref{eq:J-new}) we have
\begin{align}
 J_{(7)}^{(0)}\epsilon&=-i\Gamma_{(0)}~,
 &
 J_{(7)}^{(1)}\epsilon &= -i e^{i\theta_\kappa} C\left(\Gamma_{(0)}\epsilon\right)^\star~,
\end{align}
We have thus all the ingredients to explicitly evaluate the projection condition in (\ref{eqn:kappa-0}).
For the particular embedding where the D7-branes wrap $AdS_6\times S^2$, we have
\begin{align}\label{eq:Gamma0-D7}
 \Gamma_{(0)}&=\Gamma_{01234567}~,
\end{align}
where $\Gamma_a$ are the ten-dimensional Clifford algebra generators, explicit indices $0,...,5$ are frame indices on $AdS_6$ and $6,7$ are frame indices on $S^2$.
Moreover,
\begin{align}\label{eq:gammaX}
 \frac{1}{2}\gamma^{ij}X_{ij}&=\gamma^{67}X_{67}= \Gamma^{67} \tilde f_2^{-2}\mf ~,
\end{align}
where, following the notation in \cite{DHoker:2017mds}, the tilde on $f_2$ denotes that it is the radius of $S^2$ in string frame.
Finally,
\begin{align}\label{eq:detX}
 \sqrt{\det(1+X)}&=\sqrt{1+\tilde f_2^{-4}\mf^2}~.
\end{align}
Using (\ref{eq:Gamma0-D7}), (\ref{eq:gammaX}), (\ref{eq:detX}), as well as $C^2=1$ and $C\Gamma^a=(\Gamma^a)^\star C$, we find
\begin{align}
 \Gamma_\kappa\epsilon&=\frac{-i}{\sqrt{\tilde f_2^4+\mf^2}}\Gamma_{01234567}\left(\tilde f_2^2\epsilon+e^{i\theta_\kappa}\mf \Gamma^{67} C^{-1}\epsilon^\star\right)~.
\end{align}
Noting that raising all indices in $\Gamma_{01234567}$ produces a sign, and using (\ref{eq:Gamma-eps}), we thus find that the projection condition (\ref{eqn:kappa-0}), 
after multiplying by $\sqrt{\tilde f_2^4+\mf ^2}$, evaluates to
\begin{align}\label{eq:BPS}
 \sum_{\eta_1\eta_2}\chi^{\eta_1\eta_2}\otimes\Big[
 \tilde f_2^2\zeta_{-\eta_1\,-\eta_2}+ie^{i\theta_\kappa}\mf \star \zeta_{-\eta_1\eta_2} -\sqrt{\tilde f_2^4+\mf ^2} \zeta_{\eta_1\eta_2}
 \Big]&=0~.
\end{align}
In order for the embedding to not break any supersymmetry, the term in square brackets has to vanish for all combinations of $\eta_1$ and $\eta_2$,
and we thus arrive at
\begin{align}
 \tilde f_2^2\zeta_{-\eta_1\,-\eta_2}+ie^{i\theta_\kappa}\mf \star \zeta_{-\eta_1\eta_2} -\sqrt{\tilde f_2^4+\mf ^2} \zeta_{\eta_1\eta_2}&=0~.
\end{align}
Using the explicit parametrization in (\ref{eq:zeta}), we immediately find that the conditions are not independent, but rather that imposing the equation to be satisfied for one combination of $\eta_1$ and $\eta_2$ implies the remaining conditions.

\subsection{Solutions}\label{sec:BPS-sol}

To solve the BPS equations (\ref{eq:BPS}), we fix $\eta_1=\eta_2=+$.
With the spinors $\zeta$ in (\ref{eq:zeta}) and $\star\zeta$ defined just above (\ref{eq:zeta}), the equation to solve becomes
\begin{align}
 \tilde f_2^2\begin{pmatrix}-\bar\alpha \\ \beta\end{pmatrix} -i e^{i\theta_\kappa}\mf \begin{pmatrix}\bar\beta \\ \alpha\end{pmatrix} - \sqrt{\tilde f_2^4+\mf ^2}\begin{pmatrix}\bar\alpha \\ \beta\end{pmatrix} &=0~.
\end{align}
We note that setting $\mf=0$ does not lead to consistent solutions unless $\alpha=0$, and we therefore assume $\mf\neq 0$ from now on.
Taking the complex conjugate of the second equation, the system we have to solve is
\begin{align}
 \left(\tilde f_2^2+\sqrt{\tilde f_2^4+\mf^2}\right)\bar\alpha+ie^{i\theta_\kappa}\mf \bar \beta&=0~,\nonumber\\
 \left(\tilde f_2^2-\sqrt{\tilde f_2^4+\mf^2}\right)\bar\beta+ie^{-i\theta_\kappa}\mf \bar \alpha&=0~.
 \label{eq:kappa-sol-0}
\end{align}
Multiplying the second equation by $(-i)e^{i\theta_\kappa}\mf^{-1} (\tilde f_2^2+\sqrt{\tilde f_2^4+\mf^2})$, which is manifestly non-zero if $\mf\neq 0$, reproduces the first equation.
The two equations are thus not linearly independent and we are left with only one complex or two real conditions.
From either of the two equations, and reality of $\tilde f_2$ and $\mf$, we conclude that $e^{-i\theta_\kappa}\bar\alpha/\bar\beta$ must be imaginary or, more explicitly,
\begin{align}\label{eq:kappa-4}
 e^{i\vartheta}=\frac{\bar\alpha\beta}{\alpha\bar\beta}&=-e^{2i\theta_\kappa}~,
\end{align}
where we recognized the combination of Killing spinor components as the phase $e^{i\vartheta}$ introduced in sec.~4.3 of \cite{DHoker:2016ujz}.
Eliminating the square root between the two equations in (\ref{eq:kappa-sol-0}) yields
\begin{align}\label{eq:kappa-3}
 2\tilde f_2^2\bar\alpha\bar\beta+i\mf \left(e^{i\theta_k}\bar\beta^2+e^{-i\theta_k}\bar\alpha^2\right)&=0~,
\end{align}
which is a real equation once (\ref{eq:kappa-4}) is satisfied. 
The BPS equations are thus (\ref{eq:kappa-4}), which determines the position of the D7 brane, and (\ref{eq:kappa-3}) which determines the flux as
\begin{align}
 \mf&=\frac{2i\tilde f_2^2\bar\alpha\bar\beta}{e^{i\theta_k}\bar\beta^2+e^{-i\theta_k}\bar\alpha^2}~.
\end{align}
To evaluate the constraint on the position of the D7-brane in (\ref{eq:kappa-4}) more explicitly, we follow through the changes of variables in eq.~(4.22) and (4.27) of \cite{DHoker:2016ujz}. This yields
\begin{align}
 e^{i\vartheta}&=\frac{e^{i\psi}-\lambda R}{1-e^{i\psi}\bar\lambda R}
 =\frac{\bar\cL-\lambda\cL R}{\cL-\bar\lambda R\bar \cL}~,
\end{align}
where  we used that $e^{i\psi}=\bar \cL/\cL$ (see (4.36) and (4.48) in \cite{DHoker:2016ujz}) to obtain the second equality.
Finally, using $\kappa_-\bar\cL=-\partial_w\cG$ as well as $\kappa_\pm=\partial_w\cA_\pm$ and $\lambda=\kappa_+/\kappa_-$
we can state the $\kappa$-symmetry condition (\ref{eq:kappa-4}) as
\begin{align}\label{eq:kappa-5}
 e^{i\vartheta}&=\frac{\partial_{\bar w}\bar\cA_-\partial_w\cG-R\partial_w\cA_+\partial_{\bar w}\cG}{\partial_{w}\cA_-\partial_{\bar w}\cG-R\partial_{\bar w}\bar\cA_+\partial_{w}\cG}
 \stackrel{!}=-\frac{1+i\bar\tau}{1-i\tau}=-e^{2i\theta_\kappa}~.
\end{align}
This is one real condition on the complex position of the D7-brane in $\Sigma$, and we thus expect a one-parameter family of solutions.
We may evaluate this condition more explicitly by using that, from the definition of $B$ as $B=(1+i\tau)/(1-i\tau)$, we have
\begin{align}
 \frac{1+i\bar\tau}{1-i\tau}&=\frac{1+B}{1+\bar B}~.
\end{align}
The condition in (\ref{eq:kappa-5}) can thus be reformulated as
\begin{align}
 \left(\partial_{\bar w}\bar\cA_-\partial_w\cG-R\partial_w\cA_+\partial_{\bar w}\cG\right)\left(B+1\right)
 +\left(\partial_{w}\cA_-\partial_{\bar w}\cG-R\partial_{\bar w}\bar\cA_+\partial_{w}\cG\right)\left(\bar B+1\right)&=0~.
\end{align}
Using the definition of $B$ in (\ref{eq:B}) as well as the explicit expressions for $\partial_w\cG$ and $\partial_{\bar w}\cG$ in terms of $\cA_\pm$ and $\partial_w\cA_\pm$ that follow from the definitions in (\ref{eq:kappa-G}), this evaluates to
\begin{align}\label{eq:kappa-6}
 (1+R)\kappa^2\left(\cA_++\bar\cA_+-\cA_--\bar\cA_-\right)&=0~.
\end{align}
Since $R\geq 0$ the first factor does not vanish. For $\kappa^2\rightarrow 0$, the denominators in the original equation (\ref{eq:kappa-5}), by which we have multiplied, vanish, and a more careful treatment is needed. It shows that $\kappa^2=0$ is actually not a solution. This leaves the case where the combination of $\cA_\pm$ and their conjugates in the last factor of (\ref{eq:kappa-6}) has to vanish.
The latter condition, using $\bar\cA_\pm^0=-\cA_\mp^0$ and $\bar Z^\ell_\pm=-Z^\ell_\mp$, evaluates to
\begin{align}\label{eq:BPS-w}
 2\cA_+^0-2\cA_-^0+\sum_{\ell=1}^L(Z_+^\ell-Z_-^\ell)\ln|w-p_\ell|^2&=0~.
\end{align}
This is our final form for the $\kappa$-symmetry condition restricting the position of the probe D7-brane.
We discuss the field equations derived from the DBI action with Wess-Zumino terms in app.~\ref{app:bps-vs-field}, and have verified for several explicit examples that the BPS equations imply the field equations.

\subsection{Relation to backreacted solutions}
The BPS condition for the probe D7-brane in (\ref{eq:BPS-w}) can be directly related to the regularity conditions for the warped $AdS_6$ solutions with monodromy in (\ref{eq:w1-summary}) and (\ref{eq:DeltaG0-summary}). The regularity conditions in (\ref{eq:w1-summary}) and (\ref{eq:DeltaG0-summary}) constrain the parameters for solutions with an arbitrary number of punctures and relative weights $n_i$, and in particular also for the case that we consider one puncture with $n\equiv n_1$ infinitesimally small. To recover the probe analysis, we 
take the residues of the seed solution, $Z_\pm^\ell$, as given (with the constraint that they sum to zero) and determine the remaining parameters as formal power series in $n$
from the regularity conditions. The ansatz for the parameters is
\begin{align}
 \cA_\pm^0&=\cA_{\pm,0}^0+n^2\cA_{\pm,2}^0+\dots~,
 &
 p_\ell&=p_{\ell,0}+n^2p_{\ell,2}+\dots~,
 \nonumber\\
 w_i&=w_{i,0}+n^2w_{i,2}+\dots~.
\end{align}
At zeroth order in $n$, the conditions in (\ref{eq:DeltaG0-summary}) then reduce to the regularity conditions for a solution without monodromy, as given in (\ref{eqn:constr}). The conditions in 
(\ref{eq:w1-summary}), on the other hand, reduce precisely to the form of the $\kappa$-symmetry condition in (\ref{eq:BPS-w}). This independently supports the identification of the punctures with 7-branes.

\section{\texorpdfstring{$S^5$}{S5} partition function with backreacted 7-branes}\label{sec:backreacted}

In this section we turn to solutions with fully backreacted 7-branes and study the sphere partition functions of the dual SCFTs.
We will focus on a class of 3-pole solutions and a class of 4-pole solutions. Implications for the relation to 5-brane webs will be discussed in sec.~\ref{sec:brane-web}.

\subsection{Sphere partition function}

We now compute the sphere partition functions of the dual SCFTs for the class of 3-pole solutions illustrated in fig.~\ref{fig:3-pole-disc}. Since the SCFT is defined in odd dimensions, the renormalized sphere partition function is expected to be equal, up to a sign, to the finite part of the entanglement entropy for a ball-shaped region \cite{Casini:2011kv}. The computation of the entanglement entropy is technically simpler and we will therefore make use of this relation. The entanglement entropy is computed holographically using the Ryu-Takayanagi prescription \cite{Ryu:2006bv}, which in our case yields
 \begin{equation}\label{eq:RT}
 S_{\rm EE}=\frac{{\rm Area}(\gamma_8)}{4 G_N}~,
 \end{equation}
where $\gamma_8$ is the co-dimension two minimal surface in the 10-dimensional bulk anchored at the boundary of Poincar\'e $AdS_6$ at the location of the entangling surface. It wraps $\Sigma$ and $S^2$ entirely. As shown in detail in \cite{Gutperle:2017tjo}, eq.~(\ref{eq:RT}) can be evaluated in terms of the data characterizing the warped $AdS_6$ solutions as
\begin{align}\label{eq:EE}
S_\mathrm{EE}&=\frac{1}{4 G_\mathrm{N}}{\rm Vol}_{S^2}\cdot\mathcal J \cdot \mathrm{Area}(\gamma_4)~,
&
\mathcal J&=\frac{8}{3} \int_\Sigma d^2 w \; \kappa^2\cG~ .
\end{align}
$\rm Area(\gamma_4)$ is the area of the co-dimension two minimal surface in $AdS_6$, anchored at the entangling surface on the conformal boundary. This area is infinite and needs to be regularized. The finite part for a ball-shaped region, which is the relevant term for computing the sphere partition function, is given by
\begin{equation}\label{eq:ball-finite}
\mathrm{Area}_\mathrm{ren}(\gamma_4)=\frac{2}{3}{\rm Vol}_{S^3}~\\.
\end{equation} 
${\rm Vol}_{S^2}$ and ${\rm Vol}_{S^3}$ are the volumes of $S^2$ and $S^3$ of unit radius, respectively, and given by ${\rm Vol}_{S^2}=4\pi$ and ${\rm Vol}_{S^3}=2\pi^2$.
Let us now focus on the properties of $\cJ$. There are two different kinds of singularities that could potentially affect the evaluation of the integral: the presence of poles on $\partial \Sigma$ and the punctures inside $\Sigma$. As shown in \cite{Gutperle:2017tjo}, the integrand of $\mathcal{J}$ close to a pole behaves as $\mathcal{O}(r| \ln r|)$, with $r$ a radial coordinate centered on the pole, which is integrable.
This leaves the puncture. As shown in \cite{DHoker:2017zwj}, the asymptotic behavior of $\kappa^2$ and $\cG$ near the puncture is given by
\begin{align}
\kappa^2&\approx \mathcal{O}(\ln r)~, & \cG&\approx\mathcal{O}(1)~,
\end{align}
where $r$ is again a radial coordinate centered on the puncture. With an additional factor of $r$ coming from the measure of integration in radial coordinates, the integrand near the puncture behaves as $\mathcal{O}(r| \ln r|)$, and is again integrable.
We note that both, $\cG$ and $\kappa^2$, are single-valued functions. In addition, they both vanish at the boundary. We can thus use the relation $\kappa^2=-\partial_{\bar{w}} \partial_w \cG$ and integration by parts to obtain
\begin{equation}\label{eq:cJ}
\mathcal J=\frac{8}{3} \int_\Sigma d^2w\, |\partial_w \cG|^2~.
\end{equation}
This saves an extra integration to obtain $\cG$ and can thus be evaluated more efficiently.

The behavior of the entanglement entropy and thus the partition function under overall rescalings of the charges can be obtained from a general scaling analysis, similar to the one carried out in sec.~3 of \cite{Gutperle:2017tjo} for solutions without monodromy. The new aspect here of course is the presence of the punctures. From the explicit expression in (\ref{eqn:cA-monodromy}), one can see that $\cA_\pm$ transform homogeneously under the following rescaling of the charges
\begin{align}
 Z_\pm^\ell &\rightarrow a Z_\pm^\ell~, &n_i&\rightarrow n_i~, & a\in\RR~.
\end{align}
That is, the 5-brane charges are rescaled but the 7-brane monodromies are unchanged.   The regularity conditions in (\ref{eq:w1-summary}), (\ref{eq:DeltaG0-summary}) are invariant if $\cA^0_\pm\rightarrow a\cA_\pm^0$ with the $p_\ell$ and $w_i$ unchanged. We thus find a solution again but with $\cA_\pm\rightarrow a\cA_\pm$. This implies $\partial_w\cG\rightarrow a^2\partial_w\cG$ and thus
\begin{align}\label{eq:SEE-scaling}
 S_{\rm EE}&\rightarrow |a|^4 S_{\rm EE}~.
\end{align}
This scaling in particular holds for a ball-shaped region and therefore also applies for the sphere partition function. For the case of no punctures this reduces to the scaling derived in \cite{Gutperle:2017tjo}. The punctures therefore do not alter the scaling behavior, provided that they are not scaled with the 5-brane charges.

\subsection{Dependence on branch cut orientation}\label{sec:branch-cut-dependence}

In this section we will show that, for generic solutions, the partition function is invariant under changes of the orientation of the branch cut, as long as no poles are crossed.
More specifically, we will establish two results. The first one is that varying the orientation of the branch cut with fixed $Z_+^\ell$ does not change the partition function. Keeping the $Z_+^\ell$ fixed, however, means that the actual 5-brane charges at the poles, given by the $\cY_+^\ell$ via (\ref{eq:5-brane-charge-Y}), change. The second result shows that this change amounts to an overall $SL(2,\RR)$ transformation, which leaves the puncture and the 7-brane charge invariant. One may therefore compensate it with the inverse $SL(2,\RR)$ transformation, under which the partition function is, again, invariant. Together these results imply that the partition function is invariant under changes of the orientation of the branch cut with fixed charges of the external 5-branes, $\cY_+^\ell$.

To show that the partition function is invariant under changes of the branch cut orientation for fixed $Z_+^\ell$, 
we set up an infinitesimal shift of one of the $\gamma_i$ as follows,
\begin{align}
 \gamma_i&\rightarrow \gamma_i(1+i\delta \gamma)~.
\end{align}
Since $\gamma_i$ is a phase, $\delta \gamma$ is real. Under this change, we have
\begin{align}
 f(w)&\rightarrow f(w)+\frac{in_i^2\delta\gamma}{4\pi}~.
\end{align}
The locally holomorphic functions and their differentials transform as
\begin{align}
 \partial_w\cA_\pm &\rightarrow  \partial_w\cA_\pm+\frac{in_i^2\delta\gamma}{4\pi}\left(\partial_w\cA_+-\partial_w\cA_-\right)~,
 \nonumber\\
 \cA_\pm &\rightarrow  \cA_\pm+\frac{in_i^2\delta\gamma}{4\pi}\left(\cA_+-\cA_-\right)~.
 \label{eq:delta-cA}
\end{align}
One may have allowed for an additional shift in $\cA_\pm$ of order $\delta\gamma$, which could potentially be required to solve the regularity conditions. We will now show that this transformation without extra shift is the correct one to obtain a regular solution.
The regularity conditions were given in (\ref{eq:w1-summary}), (\ref{eq:DeltaG0-summary}) and we repeat them here for convenience
\begin{align}
 0&=2\cA_+^0-2\cA_-^0+\sum_{\ell=1}^LY^\ell \ln|w_i-p_\ell|^2~,
 \label{eq:reg1}\\
 0&=2\cA_+^0\cY_-^k-2\cA_-^0\cY_+^k
 +\sum_{\ell\neq k} Z^{[\ell, k]}\ln |p_\ell-p_k|^2+Y^kJ_k~.
 \label{eq:reg2}
\end{align}
The transformation of $\cA_\pm$ implies the following change in the constant part
\begin{align}\label{eq:delta-cA0}
 \cA_\pm^0&\rightarrow \cA_\pm^0+\delta\cA^0_\pm~, &
 \delta \cA_\pm^0&=\frac{i n_i^2\delta\gamma}{4\pi}(\cA_+^0-\cA_-^0)~.
\end{align}
The last term on the right hand side in (\ref{eq:reg1}) is independent of $\gamma$, since $Y^\ell$ is defined in terms of the $Z_\pm^\ell$. The shift in the constants, $\delta\cA_\pm$ drops out in the difference $\cA^0_+-\cA^0_-$, and the equation is therefore satisfied to linear order in $\delta\gamma$.
For (\ref{eq:reg2}), we note that $J_k$, defined in (\ref{eq:Jk}), is manifestly invariant under infinitesimal changes in the orientation of the branch cut, as long as no poles are crossed. We furthermore notice that the $\cY_\pm^\ell$ change as follows,
\begin{align}\label{eq:delta-cY}
 \cY_\pm^\ell&\rightarrow  \cY_\pm^\ell +\frac{in_i^2\delta \gamma}{4\pi}Y^\ell~.
\end{align}
Together with (\ref{eq:delta-cA0}) this shows that (\ref{eq:reg2}) is also satisfied to linear order in $\delta\gamma$.
The entire change due to the shift in the orientation of the branch cut is therefore captured by (\ref{eq:delta-cA}), which may be written as an $SL(2,\RR)$ transformation
\begin{align}\label{eq:delta-gamma-SL2R}
 \cA_+&\rightarrow u \cA_+ - v \cA_-~, &u&=1+v~,
 \nonumber\\
 \cA_-&\rightarrow -\bar v\cA_+ + \bar u \cA_-~, & v&=\frac{in_i^2\delta\gamma}{4\pi}~.
\end{align}
Since $\cG$ is invariant under $SL(2,\RR)$ transformations, and the same is true for $\partial_w\cG$, the integrand in (\ref{eq:cJ}), which directly yields the partition function, is invariant under $SL(2,\RR)$. We have thus shown that the partition function is invariant under changes of the orientation of the branch cut with fixed $Z_+^\ell$, as long as no poles are crossed. 

Finally, we note that the transformation of the actual residues at the poles corresponding to the physical 5-brane charges, as given in (\ref{eq:delta-cY}), corresponds precisely to the $SL(2,\RR)$ transformation in (\ref{eq:delta-gamma-SL2R}). Performing the inverse $SL(2,\RR)$ transformation therefore yields a solution with unmodified $\cY_+^\ell$ but shifted orientation of the branch cut. In particular, the 7-brane charge is invariant under this $SL(2,\RR)$ transformation. The argument that the integrand in (\ref{eq:cJ}) is invariant under $SL(2,\RR)$ transformations again applies, and we have thus shown that the partition function is invariant under changes of the orientation of the branch cut, as long as no poles are crossed, while keeping the $\cY_+^\ell$ fixed.

\subsection{3-pole solutions with D5, NS5 and D7}\label{sec:D5-NS5x2-D7}
We now turn to explicit solutions and start with a class of 3-pole solutions discussed already in \cite{DHoker:2017zwj}, where one of the external 5-brane stacks corresponds to D5 branes.
The poles and overall normalization $\sigma$ are chosen as
\begin{align}
 p_1&=1~,& p_2&=0~,&p_3&=-1~,
 &
 \sigma&=\frac{iN}{s_1}~.
\end{align}
The regularity conditions in (\ref{eq:w1-summary}) and (\ref{eq:DeltaG0-summary}) are satisfied by the choices
\begin{align}
 \cA_+^0&=iN \ln 2+\frac{1}{2}J_1~,
 &
 w_i&=i\alpha_i~,\quad \alpha_i\in\RR^+~,
\end{align}
which in particular implies $J_1=J_3$.
This solves the regularity conditions for an arbitrary number of punctures, but we will focus on the case of a single puncture with D7-brane monodromy in the following.
With $\alpha\equiv \alpha_1$ and $n\equiv n_1$ we thus have
\begin{align}
 f(w)&=\frac{n^2}{4\pi}\ln\left(\gamma\frac{w-i\alpha}{w+i\alpha}\right)~,
 &
 \alpha\in\RR^+~.
\end{align}
With $\ms=(s_1-1)/(2s_1)$, the residues are given by
\begin{align}
 \cY_+^1&=-iN\left[\ms+(\ms-\bar \ms)f(p_1)\right]~,&
 \cY_+^2&=iN~,&
 \cY_+^3&=iN\left[\ms-1+(\ms-\bar \ms)f(p_3)\right]~.
\end{align}
That is, the pole $p_2$ corresponds to D5 branes, while the charges of the other two poles depend on the position of the puncture, the orientation of the branch cut and the remaining parameters.

\begin{figure}
\centering
\subfigure[]{\label{fig:3-pole-disc}
\begin{tikzpicture}[scale=0.9]
\draw[fill=lightgray,opacity=0.2] (0,0) circle (1.5);
\draw[thick] (0,0) circle (1.5);
\draw[thick] (-1.4,0) -- (-1.6,0);
\draw[thick] (0,1.4) -- (0,1.6);
\draw[thick] (0,-1.4) -- (0,-1.6);
\node at (0.7,0.7) {$\mathbf \Sigma$};
\node at (-2.05,0) {D5};
\node at (0,1.85) {NS5};
\node at (0,-1.85) {NS5};

\draw[thick,dashed] (-0.2,0) -- (1.5,0);
\draw[thick,fill=black] (-0.2,0) circle (0.08);
\node at (-0.2,-0.4) {D7};
\end{tikzpicture}
}
\hskip 1.2in
\subfigure[]{\label{fig:4-pole-disc}
\begin{tikzpicture}[scale=0.9]
\draw[fill=lightgray,opacity=0.2] (0,0) circle (1.5);
\draw[thick] (0,0) circle (1.5);
\draw[thick] (-1.4,0) -- (-1.6,0);
\draw[thick] (1.4,0) -- (1.6,0);
\draw[thick] (0,1.4) -- (0,1.6);
\draw[thick] (0,-1.4) -- (0,-1.6);
\node at (0.7,0.7) {$\mathbf \Sigma$};
\node at (-2.05,0) {D5};
\node at (0,1.85) {NS5};
\node at (0,-1.85) {NS5};
\node at (2.05,0) {D5$^\prime$};

\draw[thick,dashed] (-0.2,0) -- (1.5,0);
\draw[thick,fill=black] (-0.2,0) circle (0.08);
\node at (-0.2,-0.4) {D7};
\end{tikzpicture}
}
\caption{On the left hand side a disc representation of the 3-pole solutions discussed in sec.~\ref{sec:D5-NS5x2-D7}. The D7 brane can be placed on the horizontal diameter of the disc. On the right hand side a disc representation of the 4-pole solutions discussed in sec.~\ref{sec:4-pole}.}
\end{figure}
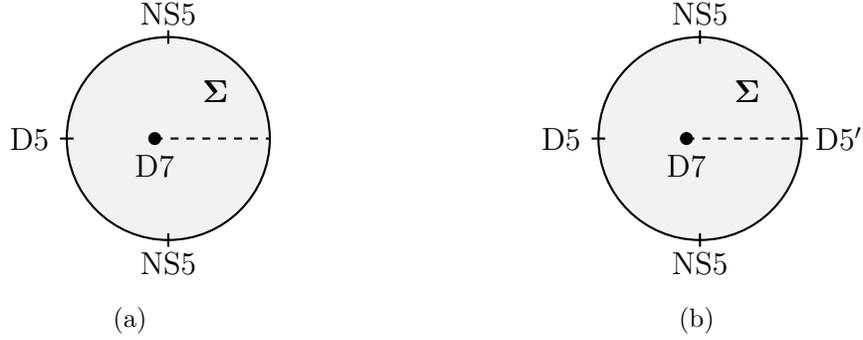

We will solve for the parameters such that the residues take the form
\begin{align}
 \cY_+^1&=M~,&
 \cY_+^2&=iN~,&
 \cY_+^3&=-M~.
\end{align}
That is, a configuration with two poles corresponding to NS5 branes, one pole corresponding to D5 branes and one puncture corresponding to D7 branes.
The setup is illustrated in fig.~\ref{fig:3-pole-disc}.
From $\cY_+^1=-\cY_+^3$ and $\cY_+^1=M$, we conclude, respectively,
\begin{align}\label{eq:cY-3pole-constr}
 (\ms-\bar \ms)\left(f(p_3)-f(p_1)\right)&=1~,
 \nonumber\\
 \ms+(\ms-\bar \ms)f(p_1)&=im~,
 &m&=\frac{M}{N}~.
\end{align}
Naively, $\cY_+^1=-\cY_+^3$ and $\cY_+^1=M$ are two complex constraints on the five remaining parameters $n$, $\alpha$, $\ms$ and $\gamma$. However, since $f$ is imaginary on the real line, the first equation in (\ref{eq:cY-3pole-constr}) is purely real, and we end up with three real constraints. We thus expect a two-parameter family of solutions. Eq.~(\ref{eq:cY-3pole-constr}) can be solved for $\ms$, which leaves only one real constraint on the parameters associated with the puncture, $n$, $\alpha$ and $\gamma$. The solution for $\ms$ and the constraint are, respectively,
\begin{align}\label{eq:D5-NS5-D7-constr}
 \ms&=im(1-2f(p_1))~,
 &
 f(p_1)-f(p_3)&=\frac{i}{2m}~.
\end{align}
We note that, with this result for $\ms$, the zero $s_1$ is in the upper half plane, as required, if and only if $m>0$.
In the following we will investigate the dependence of the sphere partition function of the dual SCFTs on the parameters associated with the puncture.

\subsubsection{Branch cut orientation}\label{sec:branch-cut-orientation}

We now discuss the orientation of the branch cut in more detail. From sec.~\ref{sec:branch-cut-dependence} we know that the partition function is independent of the choice of branch cut orientation as long as no poles are crossed. This still leaves the option for solutions with the same 5-brane and 7-brane charges, but which can not be deformed into each other without having a branch cut cross a pole.

\begin{figure}
\centering
\begin{tikzpicture}[scale=0.8]
\shade [ top color=blue! 1, bottom color=blue! 30] (2.0,1.5)  rectangle (8.0,4.5);
\node at (7.5,4.0) {$\Sigma$};

\draw[dashed,very thick,black] (5.0,3.5) -- (5.0,4.5);
\draw[black] (5.0,3.5) node {$\bullet$};
\draw (5.0,3.1) node {$w_i$};

\draw[very thick, red] plot [smooth] coordinates { (7,1.5) (6,2.2) (5,2.5) (4,2.2) (3,1.5)};

\draw [thick] (1.5,1.5) -- (8.5,1.5);
\draw (3,1.5) node{$\bullet$};
\draw (5,1.5) node{$\bullet$};
\draw (7,1.5) node{$\bullet$};
\draw (3,1.1) node{$p_{3} $};
\draw (5,1.1) node{$p_2 $};
\draw (7,1.1) node{$p_1 $};

\end{tikzpicture}
\hskip 0.8in
\begin{tikzpicture}[scale=0.8]
\shade [ top color=blue! 1, bottom color=blue! 30] (2.0,1.5)  rectangle (8.0,4.5);
\node at (7.5,4.0) {$\Sigma$};

\begin{scope}[xshift=3cm]
\draw[black] (2.0,3.5) node {$\bullet$};
\draw (2.05,3.05) node {$w_i$};
\draw[dashed,very thick,black,domain=0:1,smooth,variable=\c] plot ({2-3.4641*\c/(\c*\c+\c+1)},{3.5-4*\c*(\c+0.5)/(\c*\c+\c+1)});
\end{scope}

\draw[very thick, red] plot [smooth] coordinates { (7,1.5) (6.5,2.8) (5.8,3.8) (5,4.1) (4.2,3.8) (3.5,2.8) (3,1.5)};

\draw [thick] (1.5,1.5) -- (8.5,1.5);
\draw (3,1.5) node{$\bullet$};
\draw (5,1.5) node{$\bullet$};
\draw (7,1.5) node{$\bullet$};
\draw (3,1.1) node{$p_{3} $};
\draw (5,1.1) node{$p_2 $};
\draw (7,1.1) node{$p_1 $};

\end{tikzpicture}
\caption{
Integration contours for the constraint in (\ref{eq:D5-NS5-D7-constr-2}), depending on whether or not the branch cut intersects the boundary in the interval $(p_3,p_1)$.
\label{fig:contour}}
\end{figure}
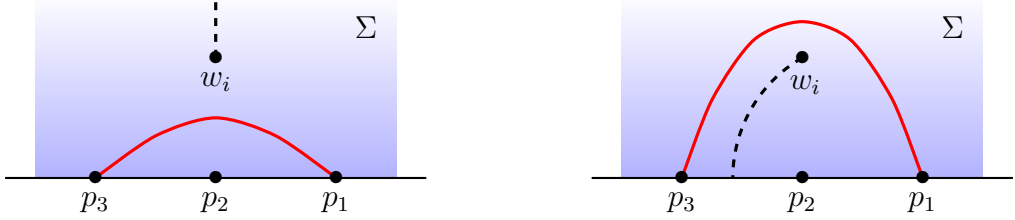

Addressing this issue requires a careful treatment of the branch cuts, and to make that explicit we rewrite the constraint on the right hand side of (\ref{eq:D5-NS5-D7-constr}) as follows
\begin{align}\label{eq:D5-NS5-D7-constr-2}
 \int_{C(p_3,p_1)}dz\, \partial_z f(z)&=\frac{i}{2m}~,
 &
 \partial_z f(z)&=\frac{i n^2}{4\pi}\frac{2\alpha}{z^2+\alpha^2}~,
\end{align}
where $C(p_3,p_1)$ denotes a contour from $p_3$ to $p_1$ that does not cross the branch cut in $f$. 
The choice of contour depends on whether the branch cut in $f$ intersects the boundary between $p_1$ and $p_3$ or not, and the choices are illustrated in fig.~\ref{fig:contour}.
If the branch cut does not intersect the boundary between $p_1$ and $p_3$, we can deform the contour to the segment of the real axis connecting $p_3$ to $p_1$ without crossing the puncture. If, on the other hand, the branch cut does intersect the boundary between $p_1$ and $p_3$, deforming the contour to the segment of the real axis between $p_3$ and $p_1$ picks up the residue at the pole $z=i\alpha$. We thus find the following constraint
\begin{align}
 \frac{i}{2m}&=-2\pi i \delta_\gamma\Res_{z=i\alpha}(\partial_z f) + \int_{p_3}^{p_1}dx f'(x) ~,
\end{align}
where we defined $\delta_\gamma=0$ if the branch cut does not intersect the boundary between $p_3$ and $p_1$, and $\delta_\gamma=1$ if it does. Evaluating the residue and the integral along the real line yields
\begin{align}
 \frac{\pi}{2mn^2}&=-\frac{\pi}{2}\delta_\gamma + \cot^{-1}\!\alpha~.
\end{align}
The left hand side is positive, in view of the fact that $m>0$ is required for $\Im(s_1)>0$.
The right hand side therefore has to be positive as well for a solution to exist. For $\alpha\in\RR^+$, however, we have $0<\cot^{-1}\!\alpha<\pi/2$. The right hand side is therefore negative if the branch cut intersects the boundary between $p_1$ and $p_3$, and the constraint can not be solved.
The remaining option is to have the branch cut intersect the boundary outside of the interval $(p_3,p_1)$, such that $\delta_\gamma=0$.
In that case a solution to the constraint exists provided that $m n^2>1$, and it is given by
\begin{align}\label{eq:D5-NS5-D7-alpha}
 \alpha&=\cot\left(\frac{\pi}{2mn^2}\right)~.
\end{align}
This solution is, in particular, independent of $\gamma$.

We thus find the following picture. Solving the regularity conditions for given 5-brane charge assignment, encoded by the $\cY_+^\ell$, and given 7-brane charge, encoded by $n^2$, imposes a `topological' constraint on the orientation of the branch cut. In the sense that it fixes between which poles the branch cut intersects the boundary, but not where exactly.

\subsubsection{Fixed orientation of the branch cut}\label{sec:gamma-fixed}

As shown in sec.~\ref{sec:branch-cut-dependence}, the partition function is invariant under changes in the orientation of the branch cut, as long as no poles are crossed, and as shown in the previous section the segment of the boundary in which the branch cut intersects intersects $\partial\Sigma$ is fixed. We now focus on the remaining dependence and keep the orientation of the branch cut, parametrized by $\gamma$, fixed.
We choose it to extend in the positive imaginary direction, such that
\begin{align}\label{eq:D5-NS5-D7}
 \gamma&=-1~,&
 s_1&=\frac{i}{2m}~.
\end{align}
This is compatible with the discussion in the previous section and the solution for $\alpha$ was given in (\ref{eq:D5-NS5-D7-alpha}).

\begin{figure}
 \centering
\subfigure[][]{\label{fig:3-pole-EE}
\begin{tikzpicture}
 \node [anchor=south west] at (0,0) {\includegraphics[width=0.42\linewidth]{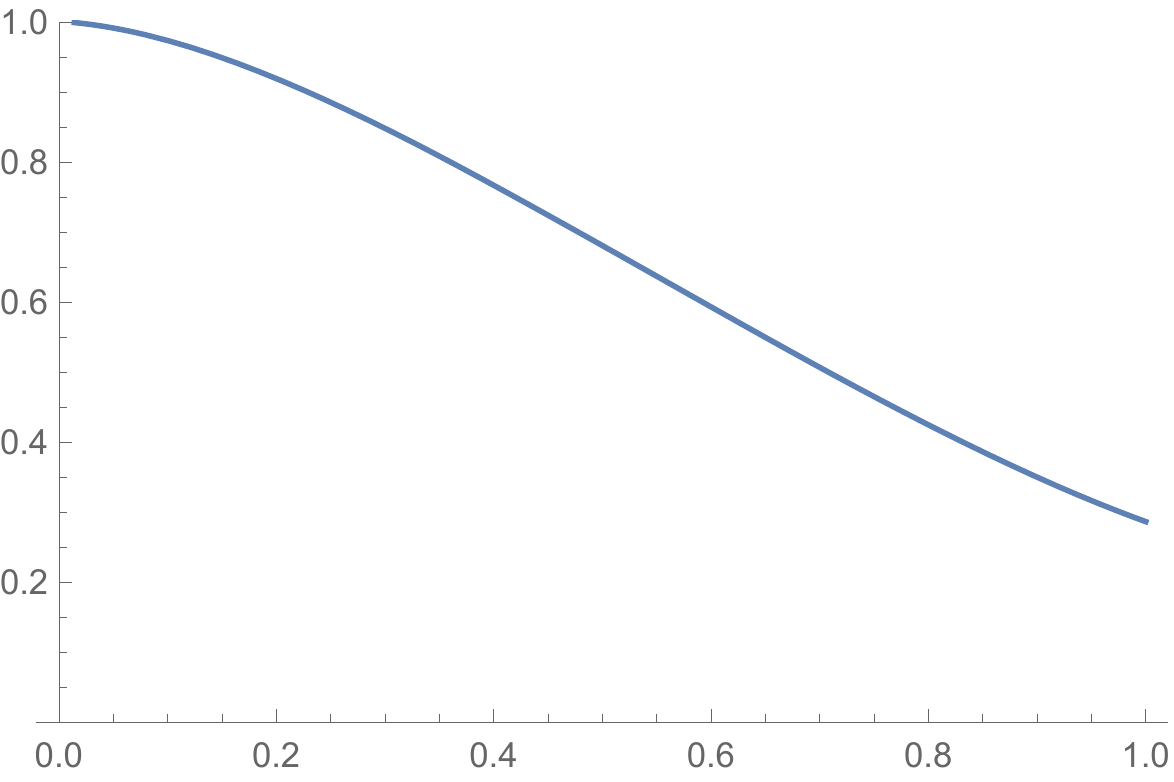}};
 \node at (6.6,0) {\small $\n$};
 \node at (4.4,3.2) {\small $\cJ_0$};
\end{tikzpicture}
}
\hfill
\subfigure[][]{\label{fig:4-pole-EE}
\begin{tikzpicture}
 \node [anchor=south west] at (0,0) {\includegraphics[width=0.42\linewidth]{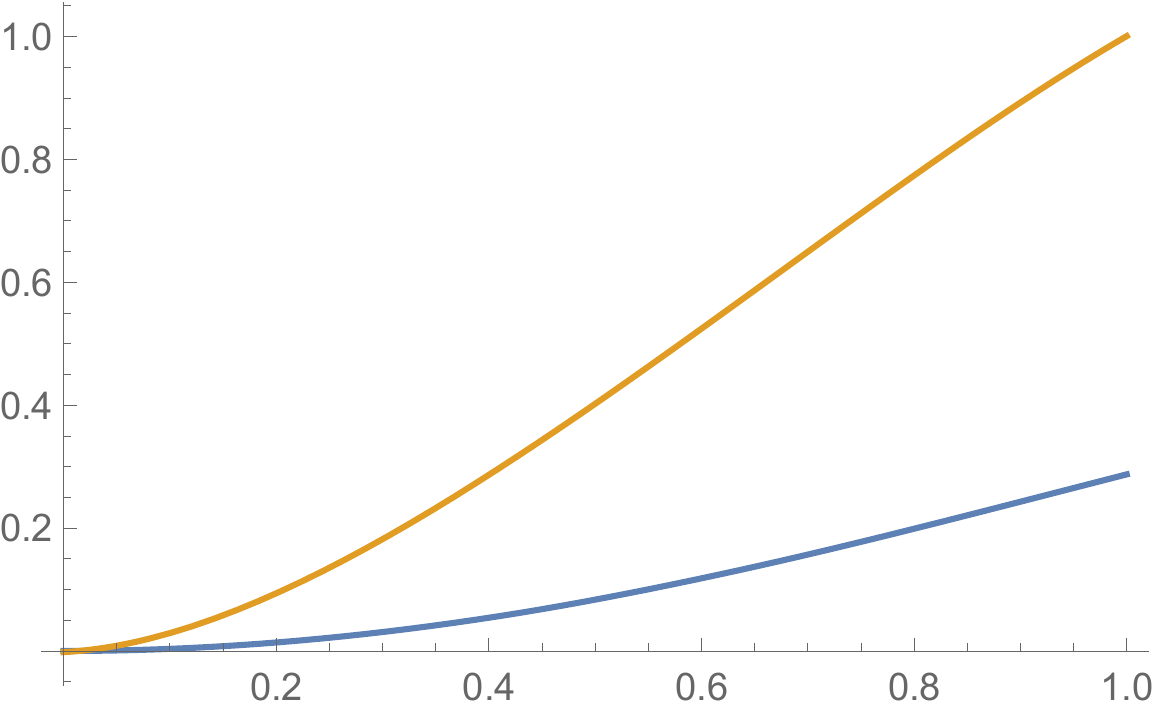}};
 \node at (6.6,0) {\small $q$};
 \node at (4.4,3.2) {\small $\cJ_{1}$};
 \node at (5.0,1.5) {\small $\cJ_{2}$};
\end{tikzpicture}
}
\caption{On the left hand side a plot of $\cJ_0$, which yields the partition function for the 3-pole solutions via (\ref{eq:cJ0}). On the right hand side similar plots for $\cJ_1$ and $\cJ_2$, which yield the partition functions of the 4-pole solutions via (\ref{eq:cJ12}).}
\end{figure}
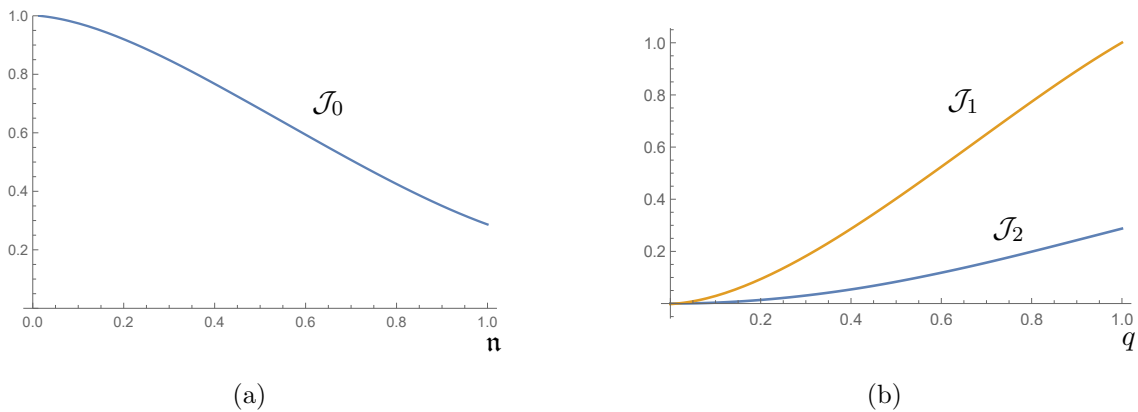

As independent parameters we take $M$, $N$ and $n^2$, while $\alpha$ is fixed by (\ref{eq:D5-NS5-D7-alpha}). To exhibit the functional dependence of the partition function, it is convenient to extract the overall scaling of the 5-brane charges. We analyze the partition function as a function of $m$ defined in (\ref{eq:cY-3pole-constr}), which is the ratio of NS5 and D5 charge, leaving $N$ as the overall scale of the 5-brane charges, and 
\begin{align}
 \n&=\frac{1}{m n^2}~,
\end{align}
which is inspired by the form of $\alpha$ in (\ref{eq:D5-NS5-D7-alpha}).
The dependence of the partition function on the overall scale of the 5-brane charges, given by $N$, is quartic, as shown in (\ref{eq:SEE-scaling}).
Since the location of the puncture depends on $\n$ only, the combination $Y^\ell f(w)$, which appears in the definition of $\cA_\pm$ and $J_k$, is independent of $m$. The $\cA_\pm$ can therefore be split into an $m$-independent part and a part linear in $m$.
Organizing the terms in $\cJ$ according to their $m$-scaling shows that only the linear part is non-vanishing, and we thus find that $\cJ$ is given by a function of $\n$ multiplied by an overall factor of $N^2M^2$. Extracting also an overall numerical factor, we parametrize it as
\begin{align}\label{eq:cJ0}
 \cJ&=224\pi\zeta(3)\,N^2 M^2 \cJ_0(\n)~. 
\end{align}
A plot of $\cJ_0(\n)$ is shown in fig.~\ref{fig:3-pole-EE}. The entanglement entropy for a ball shaped region, and thus the sphere partition function, is given by (\ref{eq:EE}) with (\ref{eq:ball-finite}) and (\ref{eq:cJ0}). The normalization in (\ref{eq:cJ0}) is chosen such that $\cJ_0=1$ reproduces the partition function of a four-pole solution without monodromy, corresponding to an intersection of D5 and NS5 branes, as discussed in \cite{Gutperle:2017tjo}.

\subsection{Turning a puncture into a pole}\label{sec:recover-4pole}
We now discuss how a 4-pole solution with D5 and NS5 branes can be recovered from the 3-pole solutions with D5 and NS5 branes and a puncture.
To this end, we start from the configuration with fixed orientation of the branch cut, as discussed in sec.~\ref{sec:gamma-fixed}.
Recall that we have three poles at
\begin{align}\label{eq:3-pole-poles}
 p_1&=1~, & p_2&=0~, & p_3&=-1~,
\end{align}
with residues given by
\begin{align}\label{eq:cY-3pole}
 \cY_+^1&=M~,&
 \cY_+^2&=iN~,&
 \cY_+^3&=-M~.
\end{align}
These residues could be realized by choosing the orientation of the branch cut as $\gamma=-1$, and the position of the branch cut $\alpha$ related to the number of 7-branes at the puncture, parametrized by $n^2$, as in (\ref{eq:D5-NS5-D7-alpha}), such that
\begin{align}\label{eq:f-3-pole-alpha}
f(w)&=\frac{n^2}{4\pi}\ln\left(\frac{i\alpha-w}{i\alpha+w}\right)~,
&
 \alpha&=\cot\left(\frac{\pi}{2mn^2}\right)~.
\end{align}
We will consider this family of solutions as parametrized by the location of the branch point, $\alpha$, and study the limit $\alpha\rightarrow\infty$. The relation on the right hand side in (\ref{eq:f-3-pole-alpha}) can be solved straightforwardly for $n^2$ and we can then expand for large $\alpha$, which yields
\begin{align}\label{eq:f-n-large-alpha}
 n^2&=\frac{\pi \alpha}{2m}+\mathcal O(\alpha^{-1})~, &f(w)&=\frac{iw}{4m}+\mathcal O(\alpha^{-1})~.
\end{align}
In particular, to realize a family of solution with fixed $\cY_+^\ell$ as given in (\ref{eq:cY-3pole}), the number of D7-branes at the puncture has to grow with $\alpha$ as the puncture is moved towards infinity (which is a regular point of the boundary of the disc). Due to this growing behavior, the function $f$ remains non-trivial in the limit.

We will now show that, as $\alpha\rightarrow\infty$, the differentials $\partial_w\cA_\pm$ approach those of a 4-pole solution, with the three poles on the boundary of $\Sigma$ that were present already for finite $\alpha$, and an extra pole at infinity.
The general form of the differentials for a solution with monodromy can be obtained straightforwardly from (\ref{eqn:cA-monodromy}), which yields
\begin{align}
 \partial_w\cA_\pm&=\sum_{\ell=1}^L\frac{Z_\pm^\ell}{w-p_\ell}+f(w)\sum_{\ell=1}^L\frac{Y^\ell}{w-p_\ell}~.
\end{align}
With the limiting behavior of $f$ in (\ref{eq:f-n-large-alpha}) and expressing $Z_\pm^\ell$ in terms of $\cY_\pm^\ell$ using the definition in (\ref{eq:cY}), we find
\begin{align}
 \partial_w\cA_\pm\big\vert_{\alpha\rightarrow\infty}&=\sum_{\ell=1}^L\frac{1}{w-p_\ell}\left(\cY_\pm^\ell-f(p_\ell)Y^\ell\right)
 +\frac{iw}{4m}\sum_{\ell=1}^L\frac{Y^\ell}{w-p_\ell}~.
\end{align}
For the particular family of solutions we are considering here, we have $Y^2=0$ and $Y^1=-Y^3$. 
Straightforward evaluation then shows that the terms proportional to $Y^\ell$ cancel and the differentials reduce to
\begin{align}
 \partial_w\cA_\pm\big\vert_{\alpha\rightarrow\infty}&=\sum_{\ell=1}^L\frac{\cY_\pm^\ell}{w-p_\ell}~.
\end{align}
That is, the differentials for a solution with poles at (\ref{eq:3-pole-poles}) with residues given in (\ref{eq:cY-3pole}). However, since the sum over $\cY_\pm^\ell$ does not vanish, we also have a pole at infinity, with residue given by\footnote{Solutions without monodromy and a pole at infinity have been discussed in more detail in \cite{Gutperle:2017tjo}.}
\begin{align}
\cY_\pm^4\big\vert_{\alpha\rightarrow\infty}&=-\sum_{\ell=1}^3\cY_\pm^\ell=-iN~.
\end{align}

We can thus explain the limiting behavior of the partition function computed in sec.~\ref{sec:gamma-fixed}:
As $\alpha\rightarrow\infty$, we have $\n\rightarrow 0$. As explained below (\ref{eq:cJ0}), the partition function of a four-pole solution with D5 and NS5 poles with residues $iN$ and $M$, respectively, is recovered from (\ref{eq:cJ0}) for $\cJ_0=1$. From fig.~\ref{fig:3-pole-EE} we indeed see that 
\begin{align}
 \lim_{\n\rightarrow 0}\cJ_0(\n)&=1~,
\end{align}
as we expect from the fact the the three-pole solution with puncture reduces to a four-pole solution without puncture in that limit.

\subsection{4-pole solutions with D5, NS5 and D7}\label{sec:4-pole}

In this section we discuss a class of 4-pole solutions where the physical 5-brane charges correspond to D5 and NS5 branes. We will realize the residues as follows,
\begin{align}\label{eq:4-pole-res}
 \cY_+^1&=M~, & \cY_+^2&=i N_1 & \cY_+^3&=-M & \cY_+^4&=-i N_2~.
\end{align}
That is, an intersection of D5 branes and NS5 branes, where the D5 charge is not conserved.
The setup is illustrated in fig.~\ref{fig:4-pole-disc}.
To realize these residues while keeping the expressions simple, it is convenient to move one pole off to infinity. We describe the details of this procedure in app.~\ref{sec:pole-infty}.
The regularity conditions in app.~\ref{sec:pole-infty}, with $p_4\rightarrow -\infty$, can be solved by fixing the remaining three poles as
\begin{align}
 p_1&=1~, & p_2&=0~,& p_3&=-1~,
\end{align}
and the branch point and orientation of the branch cut as
\begin{align}\label{eq:gamma-4pole}
 \gamma&=-1~, & w_i&=i\alpha_i~, \quad\alpha_i\in\RR^+~.
\end{align}
Note that this implies that the branch cut intersects the boundary of $\Sigma$ (of which the point at infinity in the upper half plane is a regular point), directly on a pole. This turns out to be of little consequence in this particular example, since the pole which is intersected by the branch cut has a purely imaginary residue. In particular, the residue $\cY_\pm^4$ is well defined and there are no subtleties in formulating the regularity conditions. We will come back to a more general discussion at the end of sec.~\ref{sec:brane-web}.
The choice in (\ref{eq:gamma-4pole}) immediately implies $f(p_1)=-f(p_3)$, and the branch point conditions (\ref{eq:w1-pL-summary}) are satisfied if $\tilde \cA_+^0=\tilde \cA_-^0$. 
Using this relation, together with $\tilde J_1=\tilde J_3$, in the remaining conditions (\ref{eq:DeltaG0-pL-summary}) shows that they reduce to just one condition fixing $\tilde \cA_+^0$ to
\begin{align}
 \tilde \cA_+^0&=\frac{1}{2}\tilde J_1-4f(p_3)M\ln 2~.
\end{align}
It remains to realize the residues (\ref{eq:4-pole-res}) by an appropriate choice of $s_1$, $s_2$ and $\sigma$, together with a relation between $\alpha$ and $n$. We choose
\begin{align}
 \alpha&=\cot\left(\frac{\pi q}{2}\right)~, & 
 q&=\frac{\Delta N}{n^2 M}~, 
 &\Delta N&=N_1-N_2~,
\label{eq:4-pole-branchpoint}
\end{align}
and the zeros $s_1$, $s_2$ as the two roots of the quadratic equation
\begin{align}\label{eq:4-pole-zeros}
 0&=N_2 s^2-2iM s-N_1~,
\end{align}
while $\tilde\sigma=iN_2$. With (\ref{eqn:residues-pL}) and (\ref{eq:cY}),
this indeed realizes the residues in (\ref{eq:4-pole-res}).
Note that we need $0<\Delta N<n^2 M$ for the branch point to be in the upper half plane. 
Moreover, for the zeros $s_1$, $s_2$ to be in the upper half plane, we need $N_2/M>0$ and $N_1 N_2>0$. In other words, $M$, $N_1$ and $N_2$ need to all have the same sign.

The quantity $\cJ$ which yields the entanglement entropy and thus the sphere partition function via (\ref{eq:EE}) can now be evaluated straightforwardly, and we parametrize it as follows
\begin{align}\label{eq:cJ12}
 \cJ&=224\pi \zeta(3) N_1^2 M^2\left(1-\cJ_{1}(q)\frac{\Delta N}{N_1}+\cJ_{2}(q)\frac{(\Delta N)^2}{N_1^2}\right)~.
\end{align}
The functions $\cJ_1(q)$ and $\cJ_2(q)$ are shown in fig.~\ref{fig:4-pole-EE}.
This class of solutions allows for some interesting limiting cases, which we will discuss now. 

For $N_1=N_2$ and $n=0$, we expect the solution to reduce to a 4-pole solution without monodromy. To realize this limit, we set $n^2=x$ and $N_1-N_2=x^2$, and then take the limit $x\rightarrow 0^+$. Taking the limit in this way ensures that the branch point moves to $+i\infty$, as can be seen from (\ref{eq:4-pole-branchpoint}). This eliminates the branch cut, as desired for recovering a solution without monodromy.
The partition function for the solution without monodromy was discussed in \cite{Gutperle:2017tjo}, and we recover it straightforwardly from (\ref{eq:cJ12}) since the term in the round brackets reduces to $1$.

For $N_2\rightarrow 0$, keeping all other parameters finite, we recover the 3-pole solution discussed in more detail in sec.~\ref{sec:D5-NS5x2-D7}, with the parameters $N_1\rightarrow N$, $q\rightarrow \n$. From eq.~(\ref{eq:4-pole-zeros}) one can see that one of the zeros moves to $+i\infty$ in this limit, annihilating the pole and thus leading back to a 3-pole solution.
The partition function has to reduce to the partition function of the 3-pole solution in that limit, which amounts to the relation
\begin{align}
 1-\cJ_1(\n)+\cJ_2(\n)&=\cJ_0(\n)~.
\end{align}
The curves shown in fig.~\ref{fig:3-pole-EE} and \ref{fig:4-pole-EE} indeed satisfy this relation.

\section{Implications for the brane web picture}\label{sec:brane-web}

As discussed in more detail in \cite{DHoker:2016ysh,DHoker:2017mds}, the $AdS_6$ solutions without monodromy have a compelling interpretation as supergravity description of 5-brane intersections. This clear interpretation is facilitated by the very natural mapping between the parameters of the supergravity solutions and the parameters fixing a 5-brane intersection: once the charges of the external 5-branes are fixed, supersymmetry completely fixes an intersection, and correspondingly a supergravity solution. 
With the introduction of punctures into the supergravity solutions and 7-branes into the 5-brane picture, this mapping of parameters becomes more involved. While there is still a clear relation of the supergravity parameters to the brane charges in the string theory picture (the 7-brane charge is given directly by $n^2$ while the physical 5-brane charges are given by the $\cY_\pm^\ell$ via (\ref{eq:5-brane-charge-Y})), the process of engineering a supergravity solution that realizes a given set of charges is more complicated. Moreover, a general analysis of the number of parameters alone is not sufficient anymore to completely specify the map between supergravity solution and brane webs.

The partition functions of the dual SCFTs may be used to discriminate different interpretations for the parameters of the supergravity solutions, since the partition functions are expected to agree for solutions that describe physically equivalent brane webs which realize the same SCFT. In the following we will discuss the mapping of parameters between supergravity solutions and brane webs, and the results on the partition functions in that context. As shown in \cite{DHoker:2017zwj}, the number of free parameters for a solution with $L$ poles and $I$ punctures is given by
\begin{align}\label{eq:params}
 2L-2+3I ~.
\end{align}
$2L-2$ parameters naturally arise as a choice of residues, $Z_+^\ell$, {\it of a seed solution}, subject to the constraint that they sum to zero. The three extra parameters per puncture correspond to the 7-brane charge, the location of the branch point on a curve in $\Sigma$, and the orientation of the branch cut. While the charge and orientation of the branch cut have a clear interpretation in the brane web picture, the freedom to choose a location on $\Sigma$ may seem puzzling. A crucial point for the interpretation of the solutions is that, upon adding punctures, the residues at the poles are modified and given by the $\cY_\pm^\ell$ in (\ref{eq:cY}) instead of $Z_\pm^\ell$, and that it is these modified residues that correspond to physical 5-brane charges. To address the interpretation of the parameters associated with the puncture, we have for that reason realized families of configurations with fixed $\cY_\pm^\ell$ in sec.~\ref{sec:D5-NS5x2-D7}.

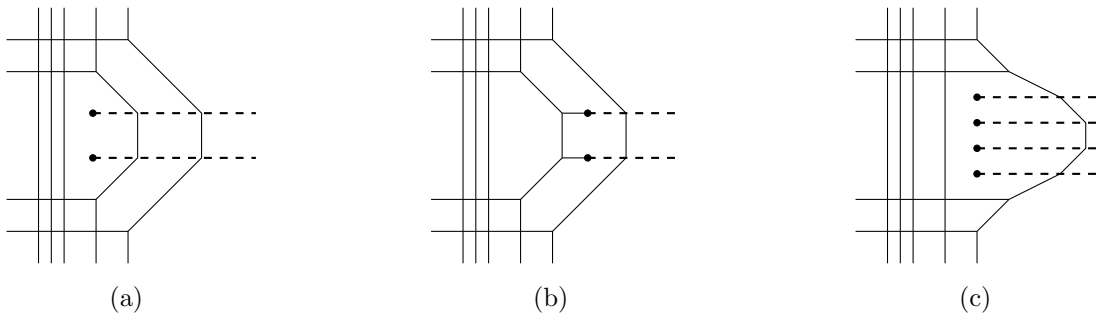
\begin{figure}
\centering
 \subfigure[][]{\label{fig:3-pole-web-1a}
 \begin{tikzpicture}[scale=0.85]
    \draw[] (-0.9,1.5) -- (1,1.5) -- (1,2);
    \draw (-0.9,1) -- (0.5,1) -- (0.5,2);
    \draw (-0.9,-1) -- (0.5,-1) -- (0.5,-2);
    \draw (-0.9,-1.5) -- (1,-1.5) -- (1,-2);
    
    \draw (0,2) -- (0,-2);
    \draw (-0.2,2) -- (-0.2,-2);
    \draw (-0.4,2) -- (-0.4,-2);
    
    \draw (0.5,1) -- (0.5+0.65,0.35) -- (0.5+0.65,-0.35) -- (0.5,-1);
    \draw (1,1.5) -- (1+1.15,0.35) -- (1+1.15,-0.35) -- (1,-1.5);
    
    \draw[fill=black] (0.45,0.35) circle (0.05);
    \draw[fill=black] (0.45,-0.35) circle (0.05);
    
    \draw[thick,dashed] (0.45,0.35) -- (3.0,0.35);
    \draw[thick,dashed] (0.45,-0.35) -- (3.0,-0.35);

 \end{tikzpicture}
 }\hskip 0.8in
 \subfigure[][]{\label{fig:3-pole-web-1b}
 \begin{tikzpicture}[scale=0.85]
    \draw[] (-0.9,1.5) -- (1,1.5) -- (1,2);
    \draw (-0.9,1) -- (0.5,1) -- (0.5,2);
    \draw (-0.9,-1) -- (0.5,-1) -- (0.5,-2);
    \draw (-0.9,-1.5) -- (1,-1.5) -- (1,-2);
    
    \draw (0,2) -- (0,-2);
    \draw (-0.2,2) -- (-0.2,-2);
    \draw (-0.4,2) -- (-0.4,-2);

    \draw (0.5,1) -- (0.5+0.65,0.35) -- (0.5+0.65,-0.35) -- (0.5,-1);
    \draw (1,1.5) -- (1+1.15,0.35) -- (1+1.15,-0.35) -- (1,-1.5);
    
    \draw[fill=black] (1.55,0.35) circle (0.05);
    \draw[fill=black] (1.55,-0.35) circle (0.05);
    
    \draw[thick,dashed] (1.55,0.35) -- (3.0,0.35);
    \draw[thick,dashed] (1.55,-0.35) -- (3.0,-0.35);
    
    \draw (1.55,-0.35) -- (1.15,-0.35);
    \draw (1.55,0.35) -- (1.15,0.35);
 \end{tikzpicture}
 }\hskip 0.8in
 \subfigure[][]{\label{fig:3-pole-web-1c}
 \begin{tikzpicture}[scale=0.85]
    
    \draw[fill=black] (1,0.6) circle (0.05);
    \draw[fill=black] (1,0.2) circle (0.05);
    \draw[fill=black] (1,-0.2) circle (0.05);
    \draw[fill=black] (1,-0.6) circle (0.05);
    \draw[thick,dashed] (1,0.6) -- (3.0,0.6);    
    \draw[thick,dashed] (1,0.2) -- (3.0,0.2);
    \draw[thick,dashed] (1,-0.2) -- (3.0,-0.2);
    \draw[thick,dashed] (1,-0.6) -- (3.0,-0.6);
    
    \draw (-0.9,-1.5) -- (1,-1.5) -- (1.5,-1) -- (2.3,-0.6) -- (2.7,-0.2) -- (2.7,0.2) -- (2.3,0.6) -- (1.5,1) -- (1,1.5) -- (-0.9,1.5);
    
    \draw (-0.9,-1) -- (1.5,-1);
    \draw (-0.9,1) -- (1.5,1);
    \draw (1,-1.5) -- (1,-2.0);
    \draw (1,1.5) -- (1,2.0);
    
    \draw (0.5,2) -- (0.5,-2);
    \draw (0,2) -- (0,-2);
    \draw (-0.2,2) -- (-0.2,-2);
    \draw (-0.4,2) -- (-0.4,-2);

 \end{tikzpicture}
 }
\caption{Fig.~\ref{fig:3-pole-web-1a} shows a possible 5-brane web corresponding to the class of supergravity solutions illustrated in fig.~\ref{fig:3-pole-disc}, with a puncture corresponding to two D7 branes. The brane web shows a general deformation of the SCFT, not the fixed point. 
Fig.~\ref{fig:3-pole-web-1b} and \ref{fig:3-pole-web-1c} show two options for 5-brane webs with the same external 5-brane charges but 7-branes in a different face of the web. The web in fig.~\ref{fig:3-pole-web-1b} is related to the web in fig.~\ref{fig:3-pole-web-1a} by 7-brane moves, the web in fig.~\ref{fig:3-pole-web-1c} is not.\label{fig:3-pole-web-moves}}
\end{figure}

In sec.~\ref{sec:gamma-fixed} we discussed the case of two NS5 brane poles, one D5 brane pole and one puncture. For fixed orientation of the branch cut {\it and} fixed $\cY_\pm^\ell$, we found a two-parameter family of solutions, where the 7-brane charge $n^2$ and the location of the puncture parametrized by $\alpha$ are related as given in (\ref{eq:D5-NS5-D7-alpha}), and the remaining parameter is the orientation of the branch cut.
Fixing a complete set of 5-brane and 7-brane charges therefore entirely fixes the configuration, up to the choice of branch cut orientation. Upon varying the position of the puncture one may keep either the 5-brane charges or the 7-brane charge fixed, but not both. This picture is consistent with the parameter count in (\ref{eq:params}) as follows. In the presence of 7-branes, the D5-brane charge is not necessarily conserved at the intersection. Fixing the 5-brane charges given by $\cY_\pm^\ell$ in the presence of punctures therefore fixes $2L-1$ parameters, instead of $2L-2$. For one puncture that leaves two free parameters, corresponding to the 7-brane charge and the orientation of the branch cut. For the case of more than one puncture, we expect relative motions of the punctures as free parameters.

To better understand the remaining parameters for one puncture, we analyzed the sphere partition function.
At fixed 5-brane and 7-brane charges, we found that the partition function does not depend on infinitesimal changes in the branch cut orientation -- at least as long as no poles are crossed. This is indeed consistent with the brane web picture: changing the orientation of the branch cut in the example web shown in fig.~\ref{fig:3-pole-web-1a}, without crossing any external 5-branes, changes the web, which describes a deformation of the SCFT. But it does not change the conformal limit, in which the web collapses to an intersection at a point. This would indeed suggest that the partition function of the UV fixed point, which is the theory described by the supergravity solution, should be independent of the precise orientation of the branch cut as long as it does not cross poles, precisely as we found in sec.~\ref{sec:branch-cut-orientation}.

The results on the partition function also allow for conclusions on the interpretation of the position of the puncture. 
We assume that the location of the puncture on $\Sigma$ corresponds to which face of the web the 7-branes are located in, which naturally becomes a continuous parameter in the ``large-$N$'' limit: with large numbers of external 5-branes, one finds a dense grid of faces, and the choice of which face the 7-branes are placed in remains meaningful in the conformal limit. One may then consider two options for supergravity solutions with the same 5-brane charges but a puncture at different positions: 
\begin{itemize}
 \item[(i)] They are related by literally moving 7-branes within the web, with the corresponding Hanany-Witten brane creation of 5-brane prongs stretching between the 7-branes and the 5-branes of the web.
 \item[(ii)] They correspond to genuinely different brane webs, where the 7-branes are placed in different faces, without 5-brane prongs stretching between the 7-branes and the 5-branes.
\end{itemize}
The two options are illustrated for a particular choice of 5-brane web in fig.~\ref{fig:3-pole-web-moves}. In case (i), one would expect the 7-brane charge to not vary as the location of the puncture is changed while keeping the external 5-brane charges fixed, as is clearly borne out by fig.~\ref{fig:3-pole-web-1b}. The field theory would remain unchanged as the location of the puncture is changed, and the same would be expected for the $S^5$ partition function of the SCFT described by the web.
In case (ii), one would expect the 7-brane charge that is required to keep the external 5-brane charges fixed to vary as the location of the puncture is varied, as is exhibited in fig.~\ref{fig:3-pole-web-1c}. The webs would describe genuinely different SCFTs and the partition functions would be expected to differ.
As we found in sec.~\ref{sec:gamma-fixed}, the charge has to be related to the location of the branch cut in a non-trivial way, as is given in (\ref{eq:D5-NS5-D7-alpha}), to preserve the external 5-brane charges. Moreover, the dependence of the partition function on the remaining free parameter is non-trivial, as can be seen explicitly from the plot in fig.~\ref{fig:3-pole-EE}. Both of these results are inconsistent with case (i), but are very well in line with option (ii). Our results show that solutions with the same 5-brane charges but punctures at different points in $\Sigma$ describe genuinely different brane webs and dual SCFTs, and the webs in fig.~\ref{fig:3-pole-web-1a} and \ref{fig:3-pole-web-1c} appear as natural brane web realizations of the solutions.

\begin{figure}
\centering
 \subfigure[][]{\label{fig:3-pole-web-1}
 \begin{tikzpicture}[scale=0.85]
    \draw[] (-0.9,1.5) -- (1,1.5) -- (1,2);
    \draw (-0.9,1) -- (0.5,1) -- (0.5,2);
    \draw (-0.9,-1) -- (0.5,-1) -- (0.5,-2);
    \draw (-0.9,-1.5) -- (1,-1.5) -- (1,-2);
    
    \draw (0,2) -- (0,-2);
    \draw (-0.2,2) -- (-0.2,-2);
    \draw (-0.4,2) -- (-0.4,-2);

    \draw (0.5,1) -- (0.5+0.65,0.35) -- (0.5+0.65,-0.35) -- (0.5,-1);
    \draw (1,1.5) -- (1+1.15,0.35) -- (1+1.15,-0.35) -- (1,-1.5);
    
    \draw[fill=black] (0.45,0.35) circle (0.05);
    \draw[fill=black] (0.45,-0.35) circle (0.05);
    
    \draw[thick,dashed] (0.45,0.35) -- (3.5,0.35);
    \draw[thick,dashed] (0.45,-0.35) -- (3.5,-0.35);

 \end{tikzpicture}
 }\hskip 1.2in
 \subfigure[][]{\label{fig:3-pole-web-2}
 \begin{tikzpicture}[scale=0.85]
    \draw[] (-0.9,1.5) -- (1,1.5) -- (1,2);
    \draw (-0.9,1) -- (0.5,1) -- (0.5,2);
    \draw (-0.9,-1) -- (0.5,-1) -- (0.5,-2);
    \draw (-0.9,-1.5) -- (1,-1.5) -- (1,-2);
    
    \draw (0,2) -- (0,-2);
    \draw (-0.2,2) -- (-0.2,-2);
    \draw (-0.4,2) -- (-0.4,-2);

    \draw (0.5,1) -- (0.5+0.69,0.31) -- (0.5+0.69,-0.31) -- (0.5,-1);
    \draw (1,1.5) -- (1+1.11,0.39) -- (1+1.11,-0.39) -- (1,-1.5);
    
    \draw[fill=black] (2.8,0.35) circle (0.05);
    \draw[fill=black] (2.8,-0.35) circle (0.05);
    
    \draw[thick,dashed] (2.8,0.35) -- (3.5,0.35);
    \draw[thick,dashed] (2.8,-0.35) -- (3.5,-0.35);
    
    \draw (2.8,0.39) -- (2.11,0.39);
    \draw (2.8,-0.39) -- (2.11,-0.39);
    
    \draw (2.8,0.31) -- (2.21,0.31);
    \draw (2.01,0.31) -- (1.19,0.31);
    \draw (2.8,-0.31) -- (2.21,-0.31);
    \draw (2.01,-0.31) -- (1.19,-0.31);

 \end{tikzpicture}
 }
\caption{
Starting from the web shown in fig.~\ref{fig:3-pole-web-1a} and moving the 7-branes out of the web along their branch cuts produces 5-brane prongs stretching between the 7-branes and the 5-branes of the web, with avoided intersections due to the $s$-rule shown as broken lines.\label{fig:3-pole-web}}
\end{figure}
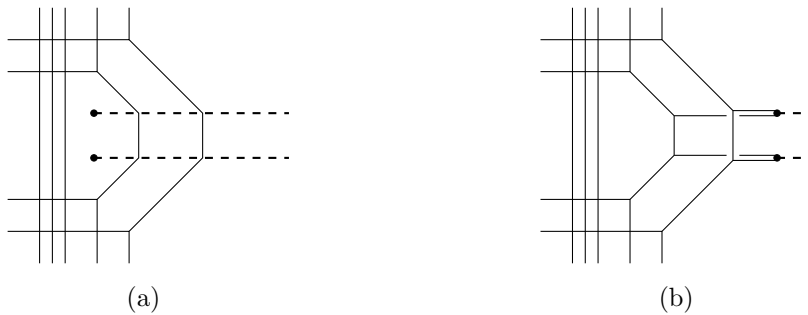

The parametrization of $\cJ$, which yields the entanglement entropy for a ball shaped region or equivalently the sphere partition function via (\ref{eq:EE}),  was chosen in (\ref{eq:cJ0}) such that $\cJ_0=1$ reproduces the partition function for a 4-pole solution with D5-brane poles and NS5-brane poles with residues $M$ and $iN$, respectively, as computed in \cite{Gutperle:2017tjo}. One might expect that a solution with 3 poles and a puncture is related to a solution with 4 poles and no puncture via Hanany-Witten transitions: pulling the D7-brane out of the 5-brane web produces a D5 brane whenever an NS5 brane is crossed, and one may suspect to get back to a solution with no puncture but an extra pole in this way. This was described in detail for an $SU(2)$ web in \cite{DeWolfe:1999hj}. However, for brane webs with large $N$ and $M$, and a D7 brane in a generic face of the web, we do not expect such a relation. The reason is illustrated in fig.~\ref{fig:3-pole-web}: due to the $s$-rule \cite{Hanany:1996ie,Benini:2009gi}, which states that no two D5 branes ending on the same 7-brane can end on the same NS5 brane while preserving supersymmetry, one would create avoided intersections in the process of pulling the D7 branes out of the web. These avoided intersections remain even if the D7-branes are moved off to infinity, and this process does therefore not lead back to a pure 5-brane web.
This explains why the partition functions for the supergravity solutions computed in sec.~\ref{sec:gamma-fixed} do in general not agree with that of a 4-pole solution without puncture.

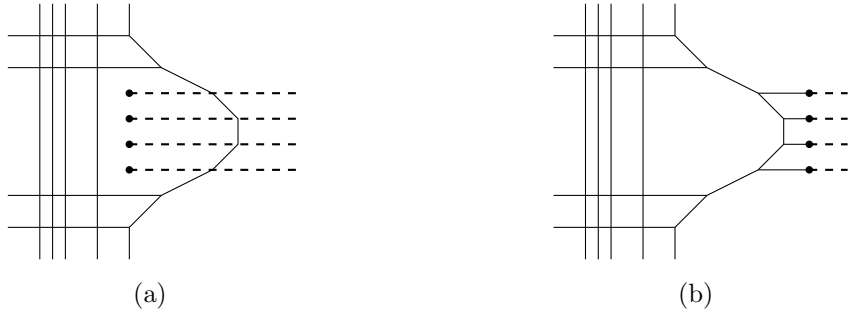
\begin{figure}
\centering
 \subfigure[][]{\label{fig:3-pole-web-3}
 \begin{tikzpicture}[scale=0.85]
    
    \draw[fill=black] (1,0.6) circle (0.05);
    \draw[fill=black] (1,0.2) circle (0.05);
    \draw[fill=black] (1,-0.2) circle (0.05);
    \draw[fill=black] (1,-0.6) circle (0.05);
    \draw[thick,dashed] (1,0.6) -- (3.7,0.6);    
    \draw[thick,dashed] (1,0.2) -- (3.7,0.2);
    \draw[thick,dashed] (1,-0.2) -- (3.7,-0.2);
    \draw[thick,dashed] (1,-0.6) -- (3.7,-0.6);
    
    \draw (-0.9,-1.5) -- (1,-1.5) -- (1.5,-1) -- (2.3,-0.6) -- (2.7,-0.2) -- (2.7,0.2) -- (2.3,0.6) -- (1.5,1) -- (1,1.5) -- (-0.9,1.5);
    
    \draw (-0.9,-1) -- (1.5,-1);
    \draw (-0.9,1) -- (1.5,1);
    \draw (1,-1.5) -- (1,-2.0);
    \draw (1,1.5) -- (1,2.0);
    
    \draw (0.5,2) -- (0.5,-2);
    \draw (0,2) -- (0,-2);
    \draw (-0.2,2) -- (-0.2,-2);
    \draw (-0.4,2) -- (-0.4,-2);
    
 \end{tikzpicture}
 }\hskip 1.2in
  \subfigure[][]{\label{fig:3-pole-web-4}
 \begin{tikzpicture}[scale=0.85]
    
    \draw[fill=black] (3.1,0.6) circle (0.05);
    \draw[fill=black] (3.1,0.2) circle (0.05);
    \draw[fill=black] (3.1,-0.2) circle (0.05);
    \draw[fill=black] (3.1,-0.6) circle (0.05);
    \draw[thick,dashed] (3.1,0.6) -- (3.7,0.6);    
    \draw[thick,dashed] (3.1,0.2) -- (3.7,0.2);
    \draw[thick,dashed] (3.1,-0.2) -- (3.7,-0.2);
    \draw[thick,dashed] (3.1,-0.6) -- (3.7,-0.6);
    
    \draw (3.1,0.6) -- (2.3,0.6);
    \draw (3.1,-0.6) -- (2.3,-0.6);
    \draw (3.1,0.2) -- (2.7,0.2);
    \draw (3.1,-0.2) -- (2.7,-0.2);
    
    \draw (-0.9,-1.5) -- (1,-1.5) -- (1.5,-1) -- (2.3,-0.6) -- (2.7,-0.2) -- (2.7,0.2) -- (2.3,0.6) -- (1.5,1) -- (1,1.5) -- (-0.9,1.5);
    
    \draw (-0.9,-1) -- (1.5,-1);
    \draw (-0.9,1) -- (1.5,1);
    \draw (1,-1.5) -- (1,-2.0);
    \draw (1,1.5) -- (1,2.0);
    
    \draw (0.5,2) -- (0.5,-2);
    \draw (0,2) -- (0,-2);
    \draw (-0.2,2) -- (-0.2,-2);
    \draw (-0.4,2) -- (-0.4,-2);    
 \end{tikzpicture}
 }

\caption{
Starting from the web shown in fig.~\ref{fig:3-pole-web-1c}, where the branch cut of each 7-brane is crossed by only one 5-brane, and moving the 7-branes out of the web, produces a pure 5-brane web with no avoided intersections.
Vertically aligning the D7-branes in the web on the right hand side with the external D5 branes turns the web into an intersection of D5 and NS5 branes. This deformation corresponds to a change of the flavor masses; the conformal UV fixed point, which is described by the supergravity solution, remains the same.
\label{fig:3-pole-web-limit}}
\end{figure}

However, as discussed in sec.~\ref{sec:recover-4pole}, we can recover a 4-pole solution without monodromy by moving the puncture along its branch cut towards the boundary of $\Sigma$, while scaling up the 7-brane charge such that the physical 5-brane charges remain invariant. This limiting procedure can be interpreted in the brane web picture as follows. For a given 7-brane we can define the notion of a distance to the ``boundary of the web'' as the number of 5-branes that cross its branch cut. For example, for the 7-branes shown in fig.~\ref{fig:3-pole-web-1a}, this distance is $2$. The limit discussed in sec.~\ref{sec:recover-4pole} can then be interpreted as increasing the number of D7-branes while placing them in faces such that their distance to the ``boundary of the web'' decreases. The transition from the web in fig.~\ref{fig:3-pole-web-1a} to the web in fig.~\ref{fig:3-pole-web-1c} gives an example of one step in this limit. The external 5-brane charges are the same for the two webs, but the distance to the ``boundary of the web'' is decreased from $2$ to $1$ in going from \ref{fig:3-pole-web-1a} to \ref{fig:3-pole-web-1c}, while the number of D7 branes is doubled.
For a supergravity solution with a puncture at a generic point on $\Sigma$, the distance to the ``boundary of the web'' of the corresponding 7-branes will be a generic number greater than one. But as the puncture is moved along its branch cut towards the boundary of $\Sigma$, this number decreases, until the 7-branes are eventually separated from the asymptotic region by only one 5-brane. Crossing this remaining 5-brane then produces 5-branes via the Hanany-Witten effect, with no constraints from the $s$-rule and no avoided intersections. For the web in fig.~\ref{fig:3-pole-web-1c} this step is shown in fig.~\ref{fig:3-pole-web-limit}. The 7-branes may now be moved off to infinity and we recover a pure 5-brane intersection. In this particular case that is an intersection of D5 and NS5 branes. This gives a brane web explanation for the fact that the partition function of a 3-pole solution with puncture agrees with the partition function of a 4-pole solution without puncture in the limit of sec.~\ref{sec:recover-4pole}.

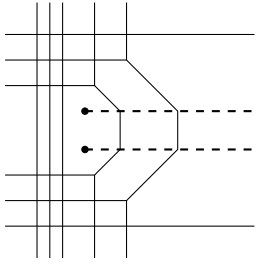
\begin{figure}
\centering
 \begin{tikzpicture}[scale=0.85]
    \draw[] (-0.9,1.1) -- (1,1.1) -- (1,2);
    \draw (-0.9,0.7) -- (0.5,0.7) -- (0.5,2);
    \draw (-0.9,-0.7) -- (0.5,-0.7) -- (0.5,-2);
    \draw (-0.9,-1.1) -- (1,-1.1) -- (1,-2);
    
    \draw (-0.9,1.5) -- (3.0,1.5);
    \draw (-0.9,-1.5) -- (3.0,-1.5);
    
    \draw (0,2) -- (0,-2);
    \draw (-0.2,2) -- (-0.2,-2);
    \draw (-0.4,2) -- (-0.4,-2);

    \draw (0.5,0.7) -- (0.5+0.4,0.3) -- (0.5+0.4,-0.3) -- (0.5,-0.7);
    \draw (1,1.1) -- (1+0.8,0.3) -- (1+0.8,-0.3) -- (1,-1.1);
    
    \draw[fill=black] (0.35,0.3) circle (0.05);
    \draw[fill=black] (0.35,-0.3) circle (0.05);
    
    \draw[thick,dashed] (0.35,0.3) -- (3.0,0.3);
    \draw[thick,dashed] (0.35,-0.3) -- (3.0,-0.3);

 \end{tikzpicture}
\caption{Brane web realization with the charges of a 4-pole solution with one puncture, as discussed in sec.~\ref{sec:4-pole}.
The number of D5 branes on the left hand side is larger than the number of D5 branes on the right hand side, corresponding to $N_1>N_2$ in the supergravity solution.
\label{fig:4-pole-web}}
\end{figure}

Finally, we studied a class of 4-pole solutions in sec.~\ref{sec:4-pole}, which provides a generalization of the 3-pole solutions. An interesting feature of these solutions is that the branch cut associated with the puncture intersects the boundary directly at the location of a pole. A natural interpretation in the brane web picture would be that the branch cut is located within a stack of external 5-branes, and an example web is shown in fig.~\ref{fig:4-pole-web}. 
In general, having a branch cut intersect a pole introduces subtleties in the supergravity description: the residue $\cY_\pm^\ell$ of the corresponding pole receives a contribution that depends on the direction from which the pole is approached (the part proportional to $f$ in (\ref{eq:cY})). While this extra contribution does not seem to obstruct the construction of regular supergravity solutions,\footnote{Since the combination of the regularity conditions associated with the remaining poles and those associated with the branch points imply the regularity condition associated with the pole intersected by the branch cut, one may simply drop the latter.} its interpretation is not entirely straightforward.
One may wonder whether it should be possible to resolve the 5-branes within a given pole in the supergravity approximation, i.e.\ whether there should be a parameter specifying between which branes exactly the branch cut is located. For the special case of a stack of D5 branes, one may move the branch cut within a given stack of branes without changing the brane web, since the charge of D5 branes does not change as they cross a branch cut associated with D7 branes. We leave a more detailed general discussion of this issue for the future, and have focused on the case with an extra D5 brane pole intersected by the branch cut in sec.~\ref{sec:4-pole}.
In this case the extra contribution to the residue drops out, since $Y^\ell$ is zero for a pole corresponding to D5 branes, and the web in fig.~\ref{fig:4-pole-web} provides a natural candidate brane web.
As discussed in sec.~\ref{sec:4-pole}, the partition function shows the correct limiting behavior in the cases where one can formulate a clear expectation for its behavior: The 3-pole solutions with puncture discussed previously can be obtained from this class of solutions as the special case where the residue of the additional D5-brane pole vanishes, while the limit where the residues at the two D5-brane poles become opposite equal leads to a solution with vanishing monodromy at the puncture and thus to a 4-pole solution without 7-branes. The partition function shows the expected behavior in these limits, but generically is a non-trivial function of the parameters. The 7-brane charge again depends non-trivially on the location of the puncture if the 5-brane charges are kept invariant, such that the results are entirely in line with the more detailed discussion of the 3-pole solutions and the conclusions drawn there.

\begin{acknowledgments}
We thank Oren Bergman and Diego Rodr{\'i}guez-G{\'o}mez for helpful discussions and comments, and 
Chrysostomos Marasinou for collaboration in the initial stage of this work.
This work is supported in part by the National Science Foundation under grant PHY-16-19926.  
AT also thanks the Mani L.\ Bhaumik Institute for summer support.
\end{acknowledgments}

\appendix

\section{Regularity conditions with a pole at infinity}\label{sec:pole-infty}

We will discuss the regularity conditions when a pole is moved to infinity. This has been discussed in \cite{Gutperle:2017tjo} for solutions without monodromy, and here we will extend this discussion to the case with monodromy.

\subsection{General reference point}
A crucial difference compared to the case without monodromy is that the construction of the holomorphic functions in (\ref{eqn:cA-monodromy}) involves a reference point to define the integral. This reference point was chosen as $+\infty$, ensuring that it does not coincidence with a pole. To keep the reference point away from poles as one of the poles is moved to infinity, the expression for $\cA_\pm$ and the regularity conditions have to be generalized to a generic reference point.
To this end, we redefine the integration constants $\cA_\pm^0$ as
\begin{align}
 \cA_\pm^0&=\hat\cA_\pm^0+\int^{\infty}_{x_0}dz f(z)\sum_{\ell=1}^L\frac{Y^\ell}{z-p_\ell}~,
\end{align}
where we assume that $x_0$ is on the real line, with no branch cuts intersecting the real line in $(x_0,\infty)$ and no poles in $(x_0,\infty)$. The expression for the holomorphic functions in (\ref{eqn:cA-monodromy}) becomes
\begin{align}\label{eqn:cA-monodromy-x0}
 \cA_\pm&= \hat\cA_\pm^0+\sum_{\ell=1}^L Z_\pm^\ell \ln(w-p_\ell) + \int_{x_0}^w dz \;f(z)\sum_{\ell=1}^L \frac{Y^\ell}{z-p_\ell}~.
\end{align}
The regularity conditions still take the same form as (\ref{eq:w1-summary}), (\ref{eq:DeltaG0-summary}), 
\begin{align}
\label{eq:w1-x0}
 0&=2\hat\cA_+^0-2\hat\cA_-^0+\sum_{\ell=1}^LY^\ell \ln|w_i-p_\ell|^2~,
 &i&=1,\cdots,I~,
 \\
 0&=2\hat\cA_+^0\cY_-^k-2\hat\cA_-^0\cY_+^k
 +\sum_{\ell\neq k} Z^{[\ell, k]}\ln |p_\ell-p_k|^2+Y^k\hat J_k~,
 &
 k&=1,\cdots,L~,
 \label{eq:DeltaG0-x0}
\end{align}
but with $\cA^0_\pm$ replaced by $\hat \cA_\pm^0$, and with $\hat J_k$ given by
\begin{align}\label{eq:Jk-x0}
 \hat J_k&=\sum_{\ell=1}^L Y^\ell\Bigg[
 f(x_0)\ln|x_0-p_\ell|^2+
 \int_{x_0}^{p_k} dx f^\prime(x)  \ln |x-p_\ell|^2
 +\sum_{i\in\cS_k} \frac{i n_i^2}{2} \ln |w_i-p_\ell|^2\Bigg]~.
\end{align}
Note that the first term in the square brackets drops out as $x_0\rightarrow\infty$ due to $\sum_\ell Y^\ell=0$.
It will be convenient to rewrite this expression for $\hat J_k$ in a more natural form, analogous to (3.46) of \cite{DHoker:2017zwj}. Namely,
\begin{align}\label{eq:Jk-x0-2}
 \hat J_k&=\sum_{\ell=1}^L Y^\ell\Big[f(x_0)\ln|x_0-p_\ell|+\int_{x_0}^{p_k}dw \ln(w-p_\ell)\partial_w f\Big]-\text{c.c.}~,
\end{align}
where the integration contour is now chosen in the upper half plane such that it does not intersect any branch cuts. This expression is obtained from (\ref{eq:Jk-x0}) by reversing the steps that were taken to get from (3.46) to (3.47) in \cite{DHoker:2017zwj}.

\subsection{Moving a pole to infinity}
We now turn to moving a pole to infinity, starting from the formulation with arbitrary reference point $x_0$.
To keep the integration constants $\hat\cA_\pm^0$ finite as the pole is moved, we combine the limit with a further redefinition of the constants in $\cA_\pm$ as follows
\begin{align}\label{eq:pole-infty}
 p_L&\rightarrow -\infty~,&
 \hat\cA_\pm^0&=\tilde \cA_\pm^0-Z_\pm^L\ln(-p_L)~,
 &
 \sigma&=-\frac{\tilde\sigma}{p_L}~,
\end{align}
where we have also redefined the overall renornalization of the residues such that the expression in (\ref{eqn:residues}) has a finite limit as $p_L\rightarrow -\infty$.
In the sum in the last term of (\ref{eqn:cA-monodromy-x0}), the $\ell=L$ term vanishes.\footnote{%
One can choose a contour such that $|z-p_L|<\kappa^{-1} |w-p_L|$ for some positive $\kappa$. 
Then $1/|z-p_L|<\kappa/|w-p_L|$ along the entire contour, and the factor can be pulled out of the integral using the Cauchy-Schwarz inequality.
The remaining integral is bounded, and the combination therefore vanishes as $p_L\rightarrow -\infty$ at fixed $w$.
} 
With the redefinition of the integration constants in (\ref{eq:pole-infty}), the expression for the locally holomorphic functions in (\ref{eqn:cA-monodromy-x0}) thus becomes
\begin{align}\label{eq:cA-infty-0}
 \cA_\pm&= \tilde\cA_\pm^0+\sum_{\ell=1}^{L-1} Z_\pm^\ell \ln(w-p_\ell) + \int_{x_0}^w dz \;f(z)\sum_{\ell=1}^{L-1} \frac{Y^\ell}{z-p_\ell}~,
\end{align}
where we have dropped terms that vanish as $p_L\rightarrow -\infty$. 
We now come to the regularity conditions.
The branch point conditions in (\ref{eq:w1-x0}) straightforwardly reduce to
\begin{align}
\label{eq:w1-pL}
 0&=2\tilde\cA_+^0-2\tilde\cA_-^0+\sum_{\ell=1}^{L-1}Y^\ell \ln|w_i-p_\ell|^2~,
 &i&=1,\cdots,I~,
\end{align}
as $p_L\rightarrow -\infty$. Of the regularity conditions in (\ref{eq:DeltaG0-x0}), we only take the subset where $k=1,\cdots,L-1$.
This was justified in the case with no punctures since only $L-1$ of the $L$ regularity conditions are independent, thanks to the fact that the residues sum to zero. In the presence of punctures, the residues do not necessarily sum to zero anymore, and the first $L-1$ conditions in (\ref{eq:DeltaG0-x0}) do {\it not} imply the last one. However, as discussed in \cite{DHoker:2017zwj}, the combination of the first $L-1$ conditions in (\ref{eq:DeltaG0-x0}) with the branch point conditions in (\ref{eq:w1-x0}) does imply the last condition in (\ref{eq:DeltaG0-x0}). This justifies dropping the $k=L$ condition and we only have to discuss the limit of
\begin{align}
 0&=2\hat\cA_+^0\cY_-^k-2\hat\cA_-^0\cY_+^k
 +\sum_{\ell\neq k} Z^{[\ell, k]}\ln |p_\ell-p_k|^2+Y^k\hat J_k~,
 &
 k&=1,\cdots,L-1~.
\end{align}
With the substitution in (\ref{eq:pole-infty}), and dropping terms that vanish as $p_L\rightarrow -\infty$, these conditions evaluate to
\begin{align}\label{eq:pL-infty-5}
 0&=2\tilde\cA_+^0\cY_-^k-2\tilde\cA_-^0\cY_+^k
 +\!\!\!\!\sum_{\ell\neq k,\ell\leq L-1}\!\!\!\! Z^{[\ell, k]}\ln |p_\ell-p_k|^2
 \nonumber\\
 &\hphantom{=}\,
 +Y^k\left[\hat J_k-2f(p_k)\ln(-p_L)Y^L\right]~.
\end{align}
It remains to evaluate the last term, noting that we only need $\hat J_k$ for $k\neq L$.
We start from (\ref{eq:Jk-x0-2}), which has the advantage that the integration contour does not cross any branch cuts.
From (\ref{eq:Jk-x0-2}) we straightforwardly find, as $p_L\rightarrow\infty$,
\begin{align}\label{eq:Jhat-pL}
 \hat J_k&=\tilde J_k+Y^L\left[f(x_0)\ln(-p_L)+\int_{x_0}^{p_k}dw \ln(-p_L)\partial_w f-\text{c.c.}\right]~,
\end{align}
where we defined
\begin{align}\label{eq:J-tilde}
 \tilde J_k&=\sum_{\ell=1}^{L-1} Y^\ell\Big[f(x_0)\ln|x_0-p_\ell|+\int_{x_0}^{p_k}dw \ln(w-p_\ell)\partial_w f\Big]-\text{c.c.}
\end{align}
The integral in the square brackets in (\ref{eq:Jhat-pL}) can be evaluated straightforwardly, and noting that $f$ is imaginary on the real axis we thus find
\begin{align}
 \hat J_k&=\tilde J_k+2Y^Lf(p_k)\ln(-p_L)~.
\end{align}
The last line in (\ref{eq:pL-infty-5}) therefore reduces to $Y^k\tilde J_k$.

In summary, we find that, for the pole $p_L$ moved to infinity, the locally holomorphic functions are given by
\begin{align}\label{eq:cA-infty-summary}
 \cA_\pm&= \tilde\cA_\pm^0+\sum_{\ell=1}^{L-1} Z_\pm^\ell \ln(w-p_\ell) + \int_{x_0}^w dz \;f(z)\sum_{\ell=1}^{L-1} \frac{Y^\ell}{z-p_\ell}~,
\end{align}
with
\begin{align}\label{eqn:residues-pL}
Z_+^\ell  &=
 \tilde\sigma\prod_{n=1}^{L-2}(p_\ell-s_n)\prod_{k \neq\ell}^{L-1}\frac{1}{p_\ell-p_k}~,
 &
 Z_-^\ell&= - \overline{Z_+^\ell}~.
\end{align}
The regularity conditions are given by
\begin{align}
\label{eq:w1-pL-summary}
 0&=2\tilde\cA_+^0-2\tilde\cA_-^0+\sum_{\ell=1}^{L-1}Y^\ell \ln|w_i-p_\ell|^2~,
 &i&=1,\cdots,I~,
 \\
 0&=2\tilde\cA_+^0\cY_-^k-2\tilde\cA_-^0\cY_+^k
 +\!\!\!\sum_{\ell\neq k,\ell\leq L-1}\!\!\! Z^{[\ell, k]}\ln |p_\ell-p_k|^2 +Y^k\tilde J_k~,
 &
 k&=1,\cdots,L-1~,
 \label{eq:DeltaG0-pL-summary}
\end{align}
with $\tilde J_k$ given by (\ref{eq:J-tilde}). The contour can be deformed to the real line, which yields
\begin{align}\label{eq:Jk-tilde}
 \tilde J_k&=\sum_{\ell=1}^{L-1} Y^\ell\Bigg[f(x_0)\ln|x_0-p_\ell|^2+\int_{x_0}^{p_k} dx f^\prime(x)  \ln |x-p_\ell|^2
 +\sum_{i\in\tilde\cS_k} \frac{i n_i^2}{2} \ln |w_i-p_\ell|^2\Bigg]~,
\end{align}
with $\tilde\cS_k$ denoting the set of poles for which the associated branch cut intersects the real line in the interval $(p_k,x_0)$.

\section{Probe BPS vs.\ field equations}\label{app:bps-vs-field}
For the warped $AdS_6$ solutions it was shown in \cite{Corbino:2017tfl} that the BPS equations imply the full set of type IIB supergravity field equations, and this in particular includes the solutions with monodromy. In this section we will derive the field equations for the probe D7 and verify that they are satisfied for configurations solving the $\kappa$-symmetry conditions of sec.~\ref{sec:BPS-sol}.

We follow the conventions of, e.g., \cite{Bergshoeff:1996tu,Simon:2011rw} for the brane effective action. 
Noting that the dilaton in the conventions used in  \cite{DHoker:2016ujz,DHoker:2016ysh,DHoker:2017mds} is related to the usual definition by a factor $2$, that results in
\begin{align}\label{eq:S-D7-gen}
 S_{D7}&=-T_7 \int d^8\xi e^{-2\phi}\sqrt{-\det\left(g_{ab}+\cF_{ab}\right)}
 +T_7\int e^\cF\wedge\sum_q C_{(q)}~,
\end{align}
where $g$ once again is the induced metric on the D7, as given in (\ref{eq:D7-metric}), and $\cF$ was defined in (\ref{eq:cF-def}).
$C_{(q)}$ denotes the RR gauge potentials of appropriate order.

The relevant RR potentials for the D7-brane are $C_{(0)}=\chi$, $C_{(2)}=\Re(\cC)\vol_{S^2}$, $C_{(6)}$ and $C_{(8)}$, where $C_{(6)}$ and $C_{(8)}$ are determined in terms of the lower RR potentials and $B_2$ \cite{Bergshoeff:2001pv}.
The symmetries of the background constrain them to take the form
 \begin{align}\label{eq:C6C8exp}
 C_{(6)}&=\mathcal C_6 \,{\rm vol}_{AdS_6} ~,
 &
 C_{(8)}&=\mathcal C_8 \, {\rm vol}_{AdS_6} \wedge {\rm vol}_{S^2} ~,
\end{align}
where $\cC_6$ and $\cC_8$ are functions on $\Sigma$. Using this parametrization as well as (\ref{mf-def}) yields
\begin{align}
 e^\cF\wedge\sum_q C_{(q)}&=\left[\mathcal C_8+\mf\,\mathcal C_6\right]\, {\rm vol}_{AdS_6} \wedge {\rm vol}_{S^2}~.
\end{align}
Moreover, the DBI part of the action can be evaluated further to yield
\begin{align}
 \int d^8\xi e^{-2\phi}\sqrt{-\det\left(g_{ab}+\cF_{ab}\right)}&=
 e^{-2\phi}\tilde f_6^6 \Vol_{AdS_6}\: \int_{S^2}\sqrt{\det \left(\tilde f_2^2 \hat g_{ S^2}+\mf{\rm vol}_{S^2}\right)}
 \nonumber\\
 &=
 e^{-2\phi}\tilde f_6^6\Vol_{AdS_6} {\rm Vol}_{S^2} \sqrt{\tilde f_2^4+\mf^2}
\end{align}
where $\hat g_{S^2}$ is the metric on $S^2$ of unit radius while $\Vol_{AdS_6}$ and $\Vol_{S^2}$ denote the (regularized) volumes of $AdS_6$ and $S^2$, respectively.
We thus find the following effective action for the D7-brane with our choice of embedding ansatz
\begin{align}\label{eq:S-D7}
 \frac{S_{D7}}{T_7 \Vol_{AdS_6}\Vol_{S^2}}&=
 -\: e^{-2\phi}\tilde f_6^6 \sqrt{\tilde f_2^4+\mf^2}
 +\mathcal C_8+\mathcal C_6\mf~.
\end{align}
We derive explicit expressions for $\cC_6$, $\cC_8$ and their field strengths in appendix \ref{sec:RR-potentials}.

We now evaluate more explicitly the equations of motion resulting from the action in (\ref{eq:S-D7}). 
We start with the worldvolume gauge field which determines $F$ via $F=dA$.
The action in (\ref{eq:S-D7-gen}) has no explicit dependence on $A$ itself, such that the equation of motion yields a conservation equation. As a result we have
\begin{align}\label{eq:F-eom}
 -e^{-2\phi}\tilde f_6^6\frac{\mf}{\sqrt{\tilde f_2^4+\mf^2}}
 +\mathcal C_6&=-\mf_0~,
\end{align}
where $\mf_0$ is a real integration constant. We emphasize that $\mf_0$ is a constant with respect to the worldvolume coordinates, but can depend on the embedding. That is, it can vary as the position of the D7-brane inside $\Sigma$ is varied.
Eq.~(\ref{eq:F-eom}) implies that $\mf$ and $\cC_6+\mf_0$ have the same sign, and the solution therefore is
\begin{align}\label{eq:eom-mf}
 \mf&=\frac{(\mathcal C_6+\mf_0) \tilde f_2^2}{\sqrt{e^{-4\phi}\tilde f_6^{12}-(\cC_6+\mf_0)^2}}~.
\end{align}
We now turn to the equation for the embedding. It can be obtained from the reduced action in (\ref{eq:S-D7}) and reads
\begin{align}
 \partial_w\Big[e^{-2\phi}\tilde f_6^6\sqrt{\tilde f_2^4+\mf^2}-\cC_8-\cC_6\mf\Big]&=0~.
\end{align}
We note that $\cK$ defined in (\ref{eq:F-0}) is to be considered as a scalar defined intrinsically on the D7-brane worldvolume. 
In particular, it does not depend on the embedding function.
We therefore have, using (\ref{mf-def}), $\partial_w\mf=-\partial_w\Re\cC$.
With the explicit expression for the R-R potentials in (\ref{eq:dC6}), (\ref{eq:dC8}), the equation of motion therefore evaluates to
\begin{align}\label{eq:eom-embedding}
 \sqrt{\tilde f_2^4+\mf^2}\partial_w\big(e^{-2\phi}\tilde f_6^6\big)+e^{-2\phi}\tilde f_6^6\frac{\partial_w\tilde f_2^4-2\mf\partial_w\Re\cC}{2\sqrt{\tilde f_2^4+\mf^2}}-i\tilde f_6^6\tilde f_2^2\partial_w\chi-\mf\partial_w\cC_6&=0~.
\end{align}
We have verified for a number of examples that solutions to the BPS equations derived in sec.~\ref{sec:BPS-sol} solve the field equations as well.
We note that the equation for the worldvolume gauge field (\ref{eq:eom-mf}) can be solved trivially by adjusting the integration constant $\mf_0$, but that the equation for the embedding in (\ref{eq:eom-embedding}) is satisfied for given $\mf$ is non-trivial.

\section{7- and 9-form R-R field strengths}\label{sec:RR-potentials}
We will describe the explicit form of the field strengths for the six- and eight-form potentials, following the conventions in \cite{Bergshoeff:2001pv}.
With $C_{(4)}=0$, the relevant field strengths $G_{(n)}$ defined in \cite{Bergshoeff:2001pv} are given by
\begin{align}
 G_{(1)}&=dC_{(0)}~,
 &
 G_{(3)}&=dC_{(2)}-C_{(0)}dB_2~,
 \\
 G_{(7)}&=dC_{(6)}~,
 &
 G_{(9)}&=dC_{(8)}-dB_2\wedge C_{(6)}~.
\end{align}
$G_{(1)}$ and $G_{(3)}$ are determined directly by the supergravity fields, while $G_{(7)}$ and $G_{(9)}$ are determined by
\begin{align}
 G_{(7)}&=-\star G_{(3)}~,
 &
 G_{(9)}&=\star G_{(1)}~,
\end{align}
where the dual is taken with respect to the string-frame metric. From these expressions we conclude
\begin{align}
 dC_{(6)}&=\star \left(C_{(0)}dB_2-dC_{(2)}\right)~,
 &
 dC_{(8)}&=\star dC_{(0)}+dB_2\wedge C_{(6)}~.
\end{align}
Using $C_{(0)}=\chi$, as well as (\ref{eq:B2-C2-cC}) and (\ref{eq:C6C8exp}), we find the more explicit expressions for the derivatives of $\cC_6$
\begin{align}
 \partial_w \cC_6&=i\tilde f_6^6\tilde f_2^{-2}\left(\chi \partial_w\Re\cC-\partial_w\Im\cC\right)~,
 \nonumber\\
 \partial_{\bar w}\cC_6&=-i\tilde f_6^6\tilde f_2^{-2}\left(\chi \partial_{\bar w}\Re\cC-\partial_{\bar w}\Im\cC\right)~,
 \label{eq:dC6}
\end{align}
and for $\cC_8$
\begin{align}
 \partial_w \cC_8&=i\tilde f_6^6\tilde f_2^{2}\partial_w\chi  +\cC_6\partial_w\Re\cC~,
 \nonumber\\
 \partial_{\bar w}\cC_8&=-i\tilde f_6^6\tilde f_2^2\partial_{\bar w}\chi+\cC_6 \partial_{\bar w}\Re\cC~.
 \label{eq:dC8}
\end{align}

\bibliographystyle{JHEP}
\bibliography{7-branes}
\end{document}